%% file: main.tex
\documentclass[aps,prx,twocolumn,superscriptaddress,showpacs,floatfix]{revtex4-2}

\usepackage{amsmath,amssymb,amsthm,amsfonts,float,graphics,epsfig,epstopdf,color,verbatim,tabularx,bm,multirow,appendix,url}
\let\oldproofname=\proofname
\renewcommand{\proofname}{\rm\bf{\oldproofname}}
\usepackage[utf8]{inputenc}
\usepackage[T1]{fontenc}
\usepackage{xcolor}
\usepackage{dsfont}
\usepackage{textcomp}
\usepackage{yfonts}
\usepackage{footnote}
\usepackage{bm}
\usepackage{subfigure}
\usepackage{mathrsfs}
\usepackage{graphicx}
\usepackage{verbatim}
\usepackage[colorlinks=true, citecolor=blue, linkcolor=blue, urlcolor=blue]{hyperref}
\usepackage{multirow}
\usepackage{enumitem,kantlipsum}
\usepackage{endnotes}

\newtheorem{theorem}{Theorem}[section]
\newtheorem{corollary}{Corollary}[theorem]

\theoremstyle{definition}
\newtheorem{definition}{Definition}[section]

\theoremstyle{remark}

\usepackage{braket}
\usepackage[normalem]{ulem}
\usepackage{tikz}
\usetikzlibrary{calc}
\usetikzlibrary{shapes.multipart}

\DeclareMathOperator{\tr}{tr}

\newcommand{\pdiff}[2]{\frac{\partial #1}{\partial #2}}

\newcommand{\be}{\begin{equation}}
\newcommand{\ee}{\end{equation}}
\newcommand{\bea}{\begin{eqnarray}}
\newcommand{\eea}{\end{eqnarray}}

\renewcommand{\vec}[1]{{\mathbf #1}}

\newcommand{\comments}[1]{}

\newcommand{\av}[1]{\langle#1\rangle}

\newcommand{\pf}[1]{\mathcal Z_{#1} }

\newcommand{\stkout}[1]{\ifmmode\text{\sout{\ensuremath{#1}}}\else\sout{#1}\fi}

\def\change#1{\textcolor{black}{#1}}


\begin{document}

\title{An analog of topological entanglement entropy for mixed states}

\author{Ting-Tung Wang}
\affiliation{Department of Physics and HK Institute of Quantum Science \& Technology, The University of Hong Kong, Pokfulam Road, Hong Kong SAR, China}

\author{Menghan Song}
\affiliation{Department of Physics and HK Institute of Quantum Science \& Technology, The University of Hong Kong, Pokfulam Road, Hong Kong SAR, China}

\author{Zi Yang Meng}
\email{zymeng@hku.hk}
\affiliation{Department of Physics and HK Institute of Quantum Science \& Technology, The University of Hong Kong, Pokfulam Road, Hong Kong SAR, China}

\author{Tarun Grover}
\email{grover@physics.ucsd.edu}
\affiliation{Department of Physics, University of California at San Diego, La Jolla, California 92093, USA}

\begin{abstract}
We propose the convex-roof extension of quantum conditional mutual information (``co(QCMI)'') as a diagnostic of \change{topological order} in a mixed state. We focus primarily on topological states subjected to local decoherence, and employ the Levin-Wen scheme to define co(QCMI), so that for a pure state, co(QCMI) equals topological entanglement entropy (TEE). By construction, co(QCMI) is zero if and only if a mixed state can be decomposed as a convex sum of pure states with zero TEE.  We show that co(QCMI) is non-increasing with increasing decoherence when Kraus operators are proportional to the product of onsite unitaries. This implies that unlike a pure state transition between a topologically trivial and a non-trivial phase, the long-range entanglement at a decoherence-induced topological phase transition as quantified by co(QCMI) is less than or equal to that in the proximate topological phase. For the  2d toric code decohered by onsite bit/phase-flip noise, we show that co(QCMI) is non-zero below the error-recovery threshold and zero above it. Relatedly, the decohered state cannot be written as a convex sum of short-range entangled pure states below the threshold. We  conjecture and provide evidence that in this example, co(QCMI) equals TEE of a recently introduced pure state. In particular, we develop a tensor-assisted Monte Carlo (TMC) computation method to efficiently evaluate the R\'enyi TEE for the aforementioned pure state and provide non-trivial consistency checks for our conjecture. We use  TMC   to also calculate the universal scaling dimension of the anyon-condensation order parameter at this transition.
\end{abstract}

\date{\today}
\maketitle

\section{Introduction} \label{sec:intro}
The entanglement structure of the ground states of local Hamiltonians has played a key role in our understanding of quantum phases of matter \cite{zeng2019quantum}. Many-body entanglement not only characterizes the universal features of ground states such as the central charge of a conformal field theory or the topological entanglement entropy (TEE) of a gapped ground state \cite{Holzhey1994,calabrese2004entanglement,hamma2005bipartite, kitaevTopological2006,levin2006detecting}, it also constrains which phases of matter or critical points can be in the vicinity of each other \cite{casini2007,casini2012,myers2010,jafferis2011towards,klebanov2011f, grover2014monotonicity}. Moreover, a coarse classification of gapped phases of matter can be argued for based on whether a ground state is short-range entangled (SRE) or long-range entangled (LRE), i.e., whether it can or cannot be obtained from a product state via a low-depth local unitary circuit \cite{verstraete2005renormalization,chen2010local,hastings2005quasiadiabatic,bravyi2006lieb}. In contrast to the ground states of local Hamiltonians, our understanding of the entanglement structure of physically realizable mixed states is relatively limited. Recent progress in defining the equivalence class of mixed states has provided a concrete definition of mixed-state phases of matter \cite{coser2019classification,sang2023mixed,sang2024stability}. It is reasonable to ask whether constraints on the entanglement structure of mixed states or their phase diagrams can be obtained based on such a definition. In this paper \change{we  propose a diagnostic of long-range entanglement in mixed states that is a close analog of TEE}. We discuss constraints imposed on this diagnostic by general considerations such as locality and renormalization group. Although the diagnostic we introduce can be defined for any mixed state, we will primarily focus on mixed states obtained by subjecting topological states such as the toric code to local decoherence \cite{dennis2002,wang2003confinement, lee2023quantum, fan2023diagnostics,bao2023mixed,wang2023intrinsic,Chen_Separability_2024,sang2023mixed,sang2024stability,li2024replica,su2024tapestry, lee2024exact,lyons2024understanding, sohal2024noisy,ellison2024towards,lu2024disentangling, kikuchi2024anyon}. For the decohered toric code, we will relate the aforementioned diagnostic to the TEE of a pure state and support our analytical arguments by calculating the TEE using a new tensor-assisted Monte Carlo (TMC)  method that integrates tensor networks with recently developed Monte Carlo algorithms for the efficient sampling of entanglement entropy~\cite{zhangIntegral2024,zhouIncremental2024}.

 \begin{figure*}[htp!]
\centering
\includegraphics[width=\textwidth]{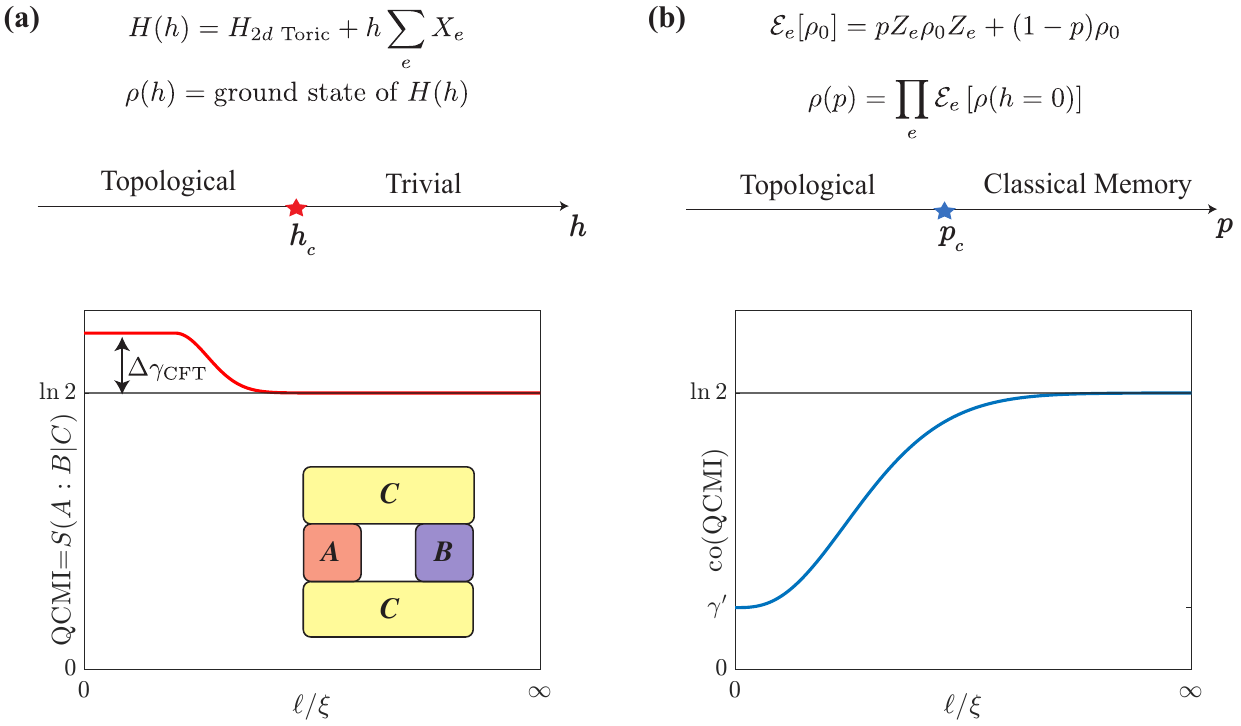}
\caption{Contrast in  entanglement scaling between pure-state transition (driven by a transverse field of strength $h$, panel (a)), and mixed-state transition (driven by decoherence at rate $p$, panel (b)) in the 2d toric code. The bottom of panel (a) schematically shows the scaling of QCMI $S(A:B|C)$ in the perturbed toric code ground state close to the topological transition when approaching  the critical point from the topological side. The diverging correlation length is denoted as $\xi$, and the linear length  of all regions $A, B, C$ that define QCMI is proportional to $\ell$. When $\ell/\xi \gg 1$, QCMI probes the topological phase, and therefore approaches TEE, i.e., $\log(2)$. On the other hand, when $\ell/\xi \ll 1$, i.e., when QCMI probes the critical regime, one receives an additional positive contribution $\Delta \gamma_{CFT}$ from the critical degrees of freedom. In contrast, when the topological order is destroyed by onsite phase-flip/bit-flip decoherence (panel (b)),  on general grounds the mixed-state entanglement captured by co(QCMI) (Eq.~\eqref{eq:def_coQCMI}) cannot exceed TEE (=$\log(2)$) in the critical regime ($ \ell/\xi \ll 1$) as discussed in Sec.~\ref{sec:coQCMIproperties} and schematically shown at the bottom of panel (b). The geometry used to define co(QCMI) is the same as the one for QCMI shown in panel (a). In fact, numerically, we find evidence that the value $\gamma'$ for the co(QCMI) in the critical regime is zero within the error-bar of our numerical simulations (Sec.~\ref{sec:numerics}, Fig.~\ref{fig:result_gamma}).}

\label{fig:CFTvsDEC}
\end{figure*}

A fundamental concept in quantum information theory is that of ``entanglement monotones'' \cite{bennett1996,vedral1997quantifying,vedral1998entanglement,vidal2000entanglement,horodecki2009quantum}. Good measures of entanglement are non-increasing under local operations, and classical communication (LOCC) operations. One may anticipate that the decrease of entanglement under LOCC operations  conforms with the naive intuition that an LRE state (e.g. a topologically ordered state) perhaps cannot be obtained from an SRE state (e.g. a trivial paramagnet without topological order) via small depth quantum channels made out of LOCC operations. However, this is not the case. The main obstacle in making such a connection is that LOCC operations only constrain the \textit{total} entanglement which includes short-distance entanglement. An LRE state can certainly be less entangled than an SRE state when short-distance entanglement is included. Indeed, by now several concrete protocols exist which allow one to prepare LRE states from SRE states via LOCC constant-depth channels \cite{briegel2001persistent,raussendorf2005long,aguado2008creation,piroli2021quantum,verresen2021efficiently,tantivasadakarn2021long,bravyi2022adaptive,lu2022shortcut,lu2023mixed}. A key idea in these protocols is to employ \textit{non-local} classical communication, which is  allowed within LOCC. In particular, the local unitaries in these protocols are contingent on the \textit{global} measurement outcomes, which is tantamount to non-local classical communication. It is therefore natural to ask what states can be obtained from a given state if \textit{only} local operations are allowed. One might anticipate that in this setting an LRE state cannot be obtained from an SRE state. If so, it is natural to ask if one can define analogs of entanglement monotones in such a setting that are sensitive only to long-range entanglement, and in a sense more universal.

Let us recall that for a bipartite Hilbert space $\mathcal{H}_A \otimes  \mathcal{H}_B$, a mixed state has zero bipartite entanglement (`separable') if it admits a representation of the form $\rho = \sum_i p_i |\psi_i \rangle \langle \psi_i|$, where each of the pure states $|\psi_i\rangle$ is bipartite unentangled, i.e., takes the form $|\psi_i\rangle = |\phi^A_i\rangle \otimes |\phi^B_i\rangle$ \cite{werner1989}. Clearly, the von Neumann entanglement $S_A$ equals zero for each $|\psi_i\rangle$. One entanglement measure that directly captures bipartite separability of a mixed state is entanglement of formation $E_F$ \cite{bennett1996}, which is defined as $E_F =  \textrm{inf} \{\sum_i p_i \,S_A(|\psi_i\rangle)\}$, where the infimum is taken over all possible pure state decompositions of the density matrix $\rho$ as $\rho = \sum_i p_i |\psi_i \rangle \langle \psi_i|$, where $p_i \geq 0$, and $\sum_i p_i = 1$. More generally, given a function $f$ from pure states to real numbers, the convex-roof extension of $f$, denoted as co$(f)$, is a function from density matrices to real numbers, and is defined as \cite{bennett1996,benatti1996optimal, vidal2000entanglement,plenio2005introduction}:
\bea 
& & \textrm{co}(f)[\rho] = \nonumber \\ 
& &  \textrm{inf} \left(\sum_i p_i \,f(|\psi_i\rangle) \Big|\,\, \rho = \sum_i p_i |\psi_i \rangle \langle \psi_i|, p_i \geq 0, \sum_i p_i = 1\right). \nonumber 
\eea 
Therefore, in this nomenclature, $E_F$ is the convex-roof extension of von Neumann entanglement \cite{bennett1996}. Our aim is to find a measure that detects whether a mixed state is SRE or not, i.e., if it admits a decomposition of the form $\sum_i p_i |\textrm{SRE}_i\rangle \langle \textrm{SRE}_i|$, where $\{|\textrm{SRE}_i\rangle \}$ are SRE pure states \cite{hastings2011topological}.  One way to achieve this is by considering the convex-roof extension of any pure state entanglement measure that captures long-range entanglement. This is the approach we will follow. Note that $E_F$ itself is not well-suited for this purpose since it will generically receive contributions from short-range entanglement (e.g. non-universal area-law contribution).

\change{For arbitrary pure states, we are unaware of a diagnostic that is non-zero if and only if the state is LRE (i.e. not obtainable from a product state via a low-depth local unitary). However, there are at least two distinct ways a pure state can be LRE which are relatively well-understood. Firstly, a state may have mutual information between distant regions that does not decay exponentially with the distance between the regions -- such LRE is archetypical for systems with spontaneous symmetry breaking (SSB), as well as ground states of gapless Hamiltonians. An example is a Greenberger-Horne-Zeilinger (GHZ) state which is representative of a system with long-range order due to SSB of $\mathbb{Z}_2$ symmetry. A different, and perhaps more profound way an LRE state can arise is due to topological order, and this kind of entanglement will be our primary focus. In a topologically ordered system, the mutual information between distant contractible regions decays exponentially, despite the fact that such states are LRE}. One way to characterize such entanglement is via topological entanglement entropy (TEE) \cite{hamma2005bipartite, kitaevTopological2006,levin2006detecting}. TEE was originally argued to equal $\log(D)$ where $D$ is the total quantum dimension for the underlying topological order. However, exceptions exist \cite{bravyi2008spurious,cano2015interactions,liujun2016spurious} (``spurious TEE'') whereby an SRE state can have non-zero Levin-Wen QCMI even when the size of the regions involved is much larger than the underlying correlation length (say, defined using connected correlators of local operators).  It was shown in Ref.~\cite{kim2023universal} that nevertheless the TEE provides a rigorous upper bound on $\log(D)$. Furthermore, in addition to the fixed-point Hamiltonians of gapped topological phases~\cite{kitaevTopological2006,levin2006detecting}, TEE equals $\log(D)$ in a variety of local Hamiltonians and variational states (see, e.g., 
Refs.~\cite{Melko_Hubbard,jiang2012identifying,zhaoMeasuring2022,zhang2011topological,nielsen2012laughlin,laflorencie2016entanglement}). 

Since a mixed state is considered short-ranged entangled if it admits a decomposition as a convex sum of short-range entangled pure states \cite{hastings2011topological}, \change{one way to partially characterize long-range entanglement of a mixed state} is to consider a measure which is zero if and only if the density matrix admits a decomposition as a convex sum of pure states with zero TEE.  This motivates us to  define the convex-roof extension of TEE as an analog of pure state TEE (Sec.~\ref{sec:coQCMIintro}).  We will define pure state TEE using Levin-Wen scheme, whereby TEE equals the quantum conditional mutual information (QCMI) for a specific choice of regions, and we will refer to the convex-roof extension of TEE as co(QCMI). By construction, if co(QCMI) is zero, then the density matrix admits a decomposition in terms of pure states with zero Levin-Wen TEE. \change{As already mentioned, zero co(QCMI) does not rule out that the mixed-state is a convex combination of pure states with GHZ-type LRE. Therefore, given a mixed-state with zero(QCMI), one may in addition also calculate its co(MI) -- the convex roof extension of mutual information -- to check whether it can be represented as a convex sum of pure states with exponentially decaying mutual information.} We will summarize some of the salient features of co(QCMI), and discuss them in detail in Sec.~\ref{sec:coQCMIproperties}. One notable feature is that for quantum channels where local Kraus operators are onsite and proportional to a unitary matrix, co(QCMI) as well as co(MI) is non-increasing. This is consistent with the expectation that an LRE mixed state cannot be obtained from an SRE mixed state via a low-depth local channel. \change{We will also discuss a generalization of this statement to channels where Kraus operators are constant-depth local unitaries but are not required to be product of onsite unitaries.} As an aside, we note that one may also define long-range part of a mixed state using convex-roof extension of pure state measures introduced in Refs.~\cite{haah2016invariant,kato2020entropic} that do not suffer from spurious TEE. We leave such explorations to the future.

 As mentioned above, one of our motivations in defining co(QCMI) is to put constraints on the phase diagrams of mixed states from mixed-state entanglement, and vice-versa. Therefore, as a testing ground, it is natural to consider co(QCMI) in models where there exist at least two different mixed-state phases as a function of some tuning parameter. A paradigmatic example is 2d or 3d toric code subjected to phase-flip and/or bit-flip noise \cite{dennis2002,wang2003confinement, lee2023quantum, fan2023diagnostics,bao2023mixed,wang2023intrinsic,Chen_Separability_2024,sang2023mixed,sang2024stability,li2024replica,su2024tapestry, lee2024exact,lyons2024understanding, sohal2024noisy,ellison2024towards,lu2024disentangling}. To setup the notation, let's consider 2d toric code whose Hamiltonian \cite{kitaev2003fault} is $H_{\textrm{2d toric}} = - \sum_v (\prod_{e \in v} Z_e) - \sum_{p} (\prod_{e \in p } X_e)$. The ground state $\rho_0$ of $H_{\textrm{2d toric}}$ is subjected to the phase-flip noise acting on an edge $e$ as: $\mathcal{E}_e[\rho_0] = p Z_e \rho_0 Z_e + (1-p) \rho_0$. The resulting phase diagram as a function of $p$ consists of two phases: for $p < p_c$, the quantum topological order survives, while for $p > p_c$, one obtains a phase with only classical topological order.  One of our main results (Sec.~\ref{sec:coQCMIproperties}) is that co(QCMI) must be non-zero for $p < p_c$, and relatedly, that the density matrix cannot be written as a convex sum of SRE pure states for $p < p_c$. This result supplements our understanding of this transition in terms of \textit{separability}: as argued in \cite{Chen_Separability_2024}, for $p > p_c$, the density matrix can be written as a convex sum of pure states that are not topologically ordered, which implies that co(QCMI) vanishes for $p > p_c$. Combined with our result, then one may view the decoherence-induced transition as a transition between a phase with non-zero co(QCMI) and a phase with zero co(QCMI). Our arguments for non-zero co(QCMI) for $p < p_c$ apply not just to 2d toric code under local decoherence, but to essentially any topologically ordered phase subjected to local decoherence. The only assumption is that the topological phase is stable for $p < p_c$ using the definition of mixed-state phase equivalence \cite{sang2023mixed,sang2024stability}, i.e., there exists a low-depth local channel that connects the decohered state to a pure topologically ordered state.

Formulating decoherence-induced transitions in terms of co(QCMI) illustrates a rather unique feature of these transitions that is not shared by quantum phase transitions in pure ground states (see Fig.~\ref{fig:CFTvsDEC}). To highlight this, consider toric code perturbed by a magnetic field: $H = H_{\textrm{2d toric}} - h \sum_e X_e$ \cite{wegner1971duality,fradkin1979phase,jongeward1980monte,tupitsyn2010topological,vidal2008low,wu2012phase,zhaoMeasuring2022}. As a function of $h$, the ground state of $H$ undergoes a phase transition from $\mathbb{Z}_2$ topologically ordered phase to a trivial paramagnet. Let's consider the bipartite entanglement entropy $S_A$ for a smooth bipartition of the total system into $A$ and $\overline{A}$ (i.e. the boundary of $A$ has no sharp corners). Let's write $S_A = \alpha \ell - \gamma + O(1/\ell)$, where $\alpha$ is a non-universal number and $\ell$ is the characteristic linear size of region $A$. In the topologically ordered phase, $\gamma = \log(2)$, while in the trivial phase, $\gamma = 0$. Right at the critical point between these two phases, $\gamma = \log(2) + \gamma_{\textrm{3d Ising}}$, where $\gamma_{\textrm{3d Ising}}$ is the subleading term in entanglement for a system described by the 2+1-d critical Ising CFT \cite{swingle2012structure,klebanov2012entanglement}. Therefore, in a sense, the critical point is more long-range entangled than either of the two phases. Indeed, due to the divergence of correlation length, the critical point is not connected to the gapped, topological phase via a low-depth local unitary. One may restate this result in terms of the behavior of the subleading term in entanglement as a function of $\ell/\xi$, where $\xi$ is the correlation length that diverges at the critical point. Approaching the critical point from the topological side, when $\ell \gg \xi$,  one finds $\gamma = \log(2)$. On the other hand, when $a \ll \ell \ll \xi$, where $a$ is the lattice spacing, one finds $\gamma = \log(2) + \gamma_{\textrm{3d Ising}}$. One may similarly consider QCMI for a Levin-Wen partition in this problem, and using the positivity of QCMI and arguments in Ref.\cite{swingle2012structure}, one then concludes that QCMI in the critical regime will equal $\log(2) + \Delta \gamma_{\text{CFT}}$, where $\Delta \gamma_{\text{CFT}}$ is the contribution from to CFT degrees of freedom, as indicated in Fig.~\ref{fig:CFTvsDEC}(a) (the magnitude of $\Delta \gamma_{\text{CFT}}$ will depend on the shape of the regions involved in defining QCMI). The fact that the critical point is more long-range entangled than the adjacent phases is a generic feature of Lorentz-invariant field theories, where one can rigorously show that the universal part of entanglement for a circle decreases under renormalization group flow \cite{ casini2007,casini2012,myers2010,jafferis2011towards,klebanov2011f}.

Now let us contrast this situation with the behavior of co(QCMI) in decoherence-induced transition in toric code (see Fig.~\ref{fig:CFTvsDEC}(b)). As already mentioned above, we will find that co(QCMI) is a non-increasing function decoherence rate, and therefore, at the critical point separating the topologically ordered mixed state to the trivial mixed state, co(QCMI) cannot be larger than $\log(2)$. Therefore, in this problem, in contrast to the pure state transition, the critical point is \textit{not} more entangled than the topological phase, at least if one uses co(QCMI) as a measure of long-range entanglement. This statement applies not just to 2d toric code under onsite decoherence, but to any onsite decoherence driven phase transition in any dimension, e.g., 3d toric code or 3d fracton models subjected to onsite phase-flip or bit-flip noise. In parallel with aforementioned discussion about pure state, let us then approach the critical point from the topological side, and probe co(QCMI) as a function of $\ell/\xi$, where $\ell$ is the characteristic size of the subsystem that defines co(QCMI) and $\xi$ is the correlation length that diverges at the critical point (defined via, say, QCMI of the corresponding mixed-state \cite{sang2024stability}). When $\ell \gg \xi$, co(QCMI) is upper-bounded by $\log(2)$. However, when $a \ll \ell \ll \xi $, in strong contrast to the pure state, our quantum many-body numerics in Sec.~\ref{sec:numerics} indicates that co(QCMI) in fact equals zero! That is, at the critical point long-range entanglement as quantified via co(QCMI) is not just smaller than that in the topological phase, it actually seems to vanish. \change{It is important to note that our result relies on the onsite nature of the decoherence, and we don't prove the monotonicity of co(QCMI) for general channels. Indeed, for general quantum channels of arbitrary depth, the entanglement of a mixed state can certainly increase (since unitaries are a subset of quantum channels, and pure states are a subset of mixed-states). Nevertheless, in Sec.\ref{sec:coQCMIproperties} we will also discuss a generalization of our result to constant-depth local mixed-unitary channels (i.e. channels where Kraus operators are proportional to constant-depth local unitaries) that are not necessarily onsite. Another point worth noting is that although  there clearly exists a low-depth local channel that takes the pure toric code to the critical point (indeed, the transition is being driven by applying such a channel), there exists no low-depth local channel in the opposite direction \cite{sang2023mixed,sang2024stability}. Intuitively, the monotonicity of the co(QCMI) originates due to the irreversible loss of quantum information to the environment. This is in contrast to the pure state transition discussed above, where there exists no   low-depth local unitary connecting the topological phase to the critical point, and vice-versa, because unitary transformations are always invertible.}


The separability-based view on the decoherence-induced phase transition in toric code naturally leads to the decomposition of the decohered toric code into a specific set of pure states that are all related to the following pure state via a local unitary transformation \cite{Chen_Separability_2024}: $\ket{\psi}=\sum_{x_\mathbf{e}}\sqrt{\pf{x_\mathbf{e}}}\ket{x_\mathbf{e}}$ where $\pf{x_\mathbf{e}}=\sum_{z_{\textbf{v}}}e^{\beta \sum_e x_e\prod_{v\in e}z_v}$ is the partition function of Ising model given bond configuration $x_{\textbf{e}}$. Ref.~\cite{Chen_Separability_2024} provided analytical arguments that the pure state $\ket{\psi}$ has zero topological entanglement entropy (TEE) for $p > p_c$, which implies that co(QCMI) is zero for $p > p_c$. The fact that the pure state $\ket{\psi}$ correctly captures the decoherence-induced phase transition, at least as far as the location and universality of the transition is considered, leads us to conjecture that the decomposition of the density matrix proposed in Ref.~\cite{Chen_Separability_2024} is optimal for co(QCMI). If this conjecture is true, then the TEE for the state $\ket{\psi}$ equals co(QCMI) of the decohered toric code, and should therefore be a monotonically non-increasing function of the decoherence rate. The underlying statistical mechanics model that enters the calculation of TEE of $\ket{\psi}$ is not exactly solvable, and the analytical arguments in Ref.~\cite{Chen_Separability_2024} do not provide information about the behavior of TEE in the vicinity of the transition.  In Sec.~\ref{sec:numerics}, we will calculate the R\'enyi TEE of the state $\ket{\psi}$ directly using state-of-the-art quantum many-body numerical methodology. To this end, we develop a tensor-assisted Monte Carlo (TMC) computational scheme to efficiently evaluate the TEE across the separability transition. The TMC method is designed to significantly mitigate the exponential complexity associated with numerical evaluation of TEE ~\cite{zhangIntegral2024,zhouIncremental2024}, and allows us to study the behavior of TEE close to the transition.

Let us summarize our main results:

\begin{enumerate}
    \item We show that the co(QCMI) for local decoherence where Kraus operators are proportional to a unitary is monotonically non-increasing as a function of the decoherence rate. Therefore, if such Kraus operators lead to a decoherence induced phase transition out of a topological phase, the co(QCMI) at the critical point cannot be larger than the TEE of the topological phase (see Fig.\ref{fig:CFTvsDEC} for an illustration in the context of 2d toric code).

    \item We show that for the 2d toric code under local phase-flip or bit-flip decoherence in any dimension, the co(QCMI) must be non-zero in the regime where error-correction is feasible. Under certain assumptions, we also argue that the value of co(QCMI) equals the TEE of the pure toric code. We also briefly discuss generalization to other topological orders.

    \item We conjecture that the decomposition of the decohered toric code introduced in Ref.~\cite{Chen_Separability_2024} is optimal for co(QCMI), and to test this conjecture, we develop a novel TMC numerical technique to compute the average TEE of this decomposition (which equals the TEE of the aforementioned state $\ket{{\psi}}$). We numerically show that the TEE of this pure state is $\log(2)$ for $p < p_c$, and zero for $p > p_c$, and that it is monotonically non-increasing as a function of the decoherence rate. These observations support our conjecture that the TEE of the state $\ket{{\psi}}$ equals the co(QCMI) of the decohered toric code.

    \item With our TMC technique, we also study the anyon condensation order parameter with respect to the aforementioned pure state $\ket{{\psi}}$. We find that the location of the transition, as well as the critical exponents match very well with the Nishimori critical point of the Random bond Ising model  \cite{nishimori1981internal,dennis2002,wang2003confinement,fan2023diagnostics,lee2023quantum}.

\end{enumerate}

\section{Convex-roof extension of quantum conditional mutual information}
\label{sec:coQCMIintro}

\subsection{\change{Brief overview of topological entanglement entropy}}
For a gapped, pure, ground state of a local Hamiltonian, topological entanglement entropy (TEE) was introduced in Refs.~\cite{levin2006detecting,kitaevTopological2006} as a diagnostic of topological order. Heuristically, for a 2d gapped ground state $|\psi\rangle$, it is defined as the subleading term $\gamma$ in the bipartite entanglement $S_A$ of a subregion $A$ of linear size $L_A$ with its complement: $S_A = \alpha L_A - \gamma$, where $\alpha$ is a non-universal constant. In 2d topologically ordered phases, the ground state consists of closed loop configurations, and intuitively, a non-zero $\gamma$ corrects for the overcounting of the entanglement between $A$ and its complement due to the closed-loop constraint on the configuration space. 

\change{Since $\gamma$ is a subleading term, and is supposed to capture the universal, long-distance physics of the gapped phase,  it is important to define it so that it is manifestly independent of the leading, non-universal contribution $\alpha L_A$. To that end, Refs.~\cite{levin2006detecting,kitaevTopological2006} introduced `subtraction schemes' to recover $\gamma$ via a combination of terms that are designed to cancel out dependence on short-distance physics. Kitaev-Preskill subtraction scheme~\cite{kitaevTopological2006} involves three regions $A, B, C$ that pairwise meet in a line-segment, and whose intersection $A \cap B \cap C$ is a point.  The TEE is obtained as $\gamma_{\text{K-P}} = \sum_i S(X_i) - \sum_{i,j} S(X_i \cup X_j) + \sum_{i,j,k} S(X_i \cup X_j \cup X_k)$, where $\{X_i \} = \{A, B, C\}$ and $S(X) = - \tr \left(\rho_X \log(\rho_X)\right)$ is the von Neumann entropy for the density matrix $\rho_X = \tr_{\overline{X}} |\psi\rangle \langle \psi|$. The Levin-Wen scheme \cite{levin2006detecting} in contrast involves the geometry `with a hole' shown in Fig.\ref{fig:CFTvsDEC}, and the TEE is  simply $\gamma_{\text{L-W}} = \frac{1}{2} S(A:B|C)$, where $S(A:B|C)  = \left( S(AC) + S(BC) - S(C) - S(ABC)\right) $ is the quantum conditional mutual information (QCMI) between $A$ and $B$ conditioned on $C$ (see Fig.~\ref{fig:tensor} (b)).  For our purposes, Levin-wen scheme will be more useful.}

We  note one subtelty about TEE. One can construct fine-tuned examples of non-topologically ordered states that have a non-zero $\gamma$ \cite{bravyi2008spurious,cano2015interactions,liujun2016spurious,kim2023universal}  (often called ``spurious TEE''). Despite this, TEE has been useful as a practical tool to detect topological order in a variety of generic models and variational wavefunctions \cite{Melko_Hubbard,jiang2012identifying,zhaoMeasuring2022,zhang2011topological,nielsen2012laughlin,laflorencie2016entanglement}. In the absence of spurious TEE,  $\gamma = \log(D)$, where $D = \sqrt{\sum_a d^2_a}$ is the total quantum dimension for the underlying topological order with  anyonic quantum dimensions $d_a$ \cite{levin2006detecting,kitaevTopological2006, kim2023universal}. One can define QCMI for any pure state, not necessarily a gapped ground state, and heuristically (assuming there is no spurious TEE), it captures one kind of ``long-range entanglement'' in a state since the combination of entropies defining it tends to cancel out the short-distance entanglement. For example, in a gapless system such as the ground state of a conformal field theory or for a system with a Fermi surface, $S(A:B|C)$ will be generically non-zero, and can be used to obtain universal data (see, e.g., \cite{faulkner2016shape}). Levin-Wen TEE however is oblivious to GHZ-type LRE as already noted in the Introduction.

\subsection{Convex room extension of TEE}
\change{Our aim is to define a quantity that is an analog of TEE for mixed states.} Unlike pure states, for which von Neumann entanglement is essentially a unique measure of bipartite entanglement, there exist several different entanglement measures for mixed states. One proposal is to define a combination analogous to $S(A:B|C)$ in the aforementioned Levin-Wen scheme by replacing each of the terms $S(AC), S(BC), S(C), S(ABC)$ with a measure of bipartite mixed-state entanglement such as negativity \cite{eisert99, vidal2002, plenio2005logarithmic, lu2020detecting,lu2023characterizing,fan2023diagnostics}. One potential issue with negativity is that it is not a faithful measure of mixed-state entanglement: there exist states that are entangled but have zero negativity. Here, we will follow a different approach by introducing a measure that is closer in spirit to TEE. We will construct a measure that is zero if and only if the mixed state admits a decomposition in terms of pure states that have zero TEE. Consider a mixed state $\rho$ over a tetrapartite Hilbert space $A\otimes B \otimes C \otimes D$, where $A, B, C$ have the same geometry as the one used to define Levin-Wen TEE, and $D$ denotes the complement of $ABC$.
\begin{definition}  \label{eq:def_coQCMI}
Given a tetra-partite density matrix $\rho_{ABCD}$, we define $\textrm{co(QCMI)}[\rho_{ABCD}] = \textrm{inf} \{\sum_i p_i \,\gamma(|\psi_i\rangle_{ABCD})\}$ where $\gamma(|\psi_i\rangle_{ABCD}) = \frac{1}{2} S(A:B|C)$ and the infimum is taken over all possible pure-state decompositions of the mixed state $\rho$ as $\rho = \sum_i p_i |\psi_i\rangle \langle \psi_i|$. 
\end{definition}
%

Thus, co(QCMI) is the convex-roof extension of $\gamma$ to mixed states, just as the entanglement of formation is the convex-roof extension of von Neumann entanglement for bipartite states \cite{bennett1996,vidal2000entanglement,horodecki2009quantum}. Due to strong subadditivity, co(QCMI)$[\rho] \geq 0$. It is worth noting that unlike entanglement of formation, co(QCMI) is generically \textit{not} an entanglement monotone under LOCC operations. This is because QCMI is neither a concave nor a convex function of density matrices \cite{vidal2000entanglement,shirokov2017tight}. Indeed, LOCC operations allow one to obtain LRE states from SRE states, via constant-depth channels due to the possibility of non-local classical communication \cite{briegel2001persistent,raussendorf2005long,aguado2008creation,piroli2021quantum,verresen2021efficiently,tantivasadakarn2021long,bravyi2022adaptive,lu2022shortcut,lu2023mixed}. However, a mixed-state phase of matter is defined via the equivalence class of states related to each other via low-depth \textit{local} channels \cite{coser2019classification,sang2023mixed,sang2024stability}, and therefore, it is desirable to seek a measure of long-range entanglement that is monotonic when only low-depth local operations are allowed.  As we will discuss in the next section, co(QCMI) is monotonic under at least a class of local low-depth channels that are of our interest.

One may also define a R\'enyi version of co(QCMI), by replacing $\gamma = \frac{1}{2}S(A:B|C)$ in Eq.\eqref{eq:def_coQCMI} by its R\'enyi version, namely, $\gamma_n = \frac{1}{2}S_n(A:B|C) =\frac{1}{2} \left( S_n(AC) + S_n(BC) - S_n(C) - S_n(ABC)\right)$, where $S_n(X) = - \frac{1}{n-1}\log \left(\tr \left(\rho^n_X\right)\right)$ is the R\'enyi entropy for the density matrix $\rho_X$ in the state $|\psi_i\rangle$. This quantity shares several features with co(QCMI) as discussed in Sec.~\ref{sec:coQCMIproperties} and is potentially more amenable to numerical simulations. We will employ it in the tensor-assisted Monte Carlo computation in  Sec.~\ref{sec:numerics}.

\section{Constraints on co(QCMI) for decoherence driven topological transitions} \label{sec:coQCMIproperties}

In this section, we will discuss some of the salient properties of co(QCMI) (Definition~\ref{eq:def_coQCMI}). Our focus will primarily be pure topologically ordered states that are being subjected to local decoherence  \cite{dennis2002,wang2003confinement, lee2023quantum, fan2023diagnostics,bao2023mixed,wang2023intrinsic,Chen_Separability_2024,sang2023mixed,sang2024stability,li2024replica,su2024tapestry, lee2024exact,lyons2024understanding, sohal2024noisy,ellison2024towards,lu2024disentangling}. A paradigmatic example is 2d toric code in the presence of phase-flip or bit-flip noise. For concreteness, we will focus on this example, and discuss along the way which features generalize. We write 2d toric code as $H_{\textrm{2d toric}} = - \sum_v (\prod_{e \in v} Z_e) - \sum_{p} (\prod_{e \in p } X_e)$. We subject a (pure) ground state $\rho_0$ of $H_{\textrm{2d toric}}$ to phase-flip channel acting on an edge $e$ as: $\mathcal{E}_e[\rho_0] = p Z_e \rho_0 Z_e + (1-p) \rho_0$ where $p \geq 0$ is the decoherence strength. The full dynamics corresponds to the composition of the map $\mathcal{E}_e[\cdot]$ on all edges, and we will  denote its action simply as $\mathcal{E}[\cdot]$. It is well known that this  system undergoes a phase transition as a function of the decoherence rate $p$ \cite{dennis2002,wang2003confinement, lee2023quantum, fan2023diagnostics,bao2023mixed}. For $p < p_c \approx 0.11$, the system retains quantum memory of the undecohered toric code ground state while the quantum memory is lost for $p > p_c$ and one enters a ``classical memory'' phase. The non-triviality of the $p < p_c$ phase can be argued from a variety of perspectives, e.g., using coherent information \cite{fan2023diagnostics,lee2024exact,wang2023intrinsic} or via mixed-state phase equivalence \cite{sang2023mixed,sang2024stability}, and the related idea of emergent anomalous one-form strong symmetries \cite{ellison2024towards}.  Since we are interested in quantifying the long-range entanglement, central to our discussion will be the separability aspects of the mixed state. On that note, Ref.~\cite{Chen_Separability_2024} argued that for $p \geq p_c$, the density matrix can be expressed as a convex sum of states that have zero TEE, which would imply that co(QCMI) is zero for $p > p_c$. It was also conjectured that such a decomposition is not possible for $p < p_c$, although a proof so far is lacking. Motivated by these considerations, let us ask a few questions:

\begin{figure}[htp!]
\centering
\includegraphics[width=\columnwidth]{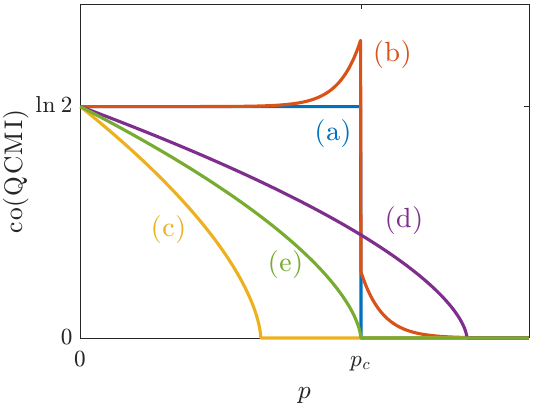}
\caption{\textbf{Possible scenarios of co(QCMI) across the decoherence transition in 2d noisy toric code.} We discuss the scenarios (a) -- (e) in addressing the questions \#1 -- \#4 below, and provide arguments that rule out scenarios (b), (c), (d) and (e).}

\label{fig:scenarios}
\end{figure}

\begin{enumerate}[wide, labelwidth=!, labelindent=0pt]
    \item Can co(QCMI) increase as the decoherence rate increases? (scenario (b) in Fig.~\ref{fig:scenarios}).
    \item Can co(QCMI) be zero for $p < p_c$? (scenario (c) in Fig.~\ref{fig:scenarios}).
    \item If the answer to question \#2 is `no',  does co(QCMI) remain quantized at $\log(2)$ for $p < p_c$? (scenario (a) in Fig.~\ref{fig:scenarios}). Or can it be less than $\log(2)$? (scenario (e) in Fig.~\ref{fig:scenarios}).
    \item Can co(QCMI) be non-zero for $p > p_c$? (scenario (d) in Fig.~\ref{fig:scenarios}).
\end{enumerate}
 We will consider these questions in turn.

     \subsection{Can co(QCMI) increase under local decoherence?} 
     
     We will start with a definition and a theorem.
   \begin{definition}\label{def:cof}
   	 Let $f$ be a real-valued function defined on the space of all pure states, i.e.,$|\psi\rangle \mapsto f(|\psi_i\rangle)$. For any density matrix $\rho$, we define
   	\be    
   	\textrm{co}(f)[\rho] = \textrm{inf}  \{\sum_i p_i \,f(|\psi_i\rangle) |\,\, \rho = \sum_i p_i |\psi_i \rangle \langle \psi_i|\}
   	\ee 
   \end{definition}  
  
We will call co$(f)$ the convex-roof of $f$.
     
     \begin{theorem} \label{theorem:cof}
     	\change{Consider convex-roof co$(f)$ of any quantity $f$ that is invariant under onsite unitary transformations, i.e., $f(|\psi\rangle)  = f(U |\psi\rangle)$, where $U = \prod_i U_i$ is a product of onsite unitaries. Consider a density matrix that is subjected to a quantum channel $\mathcal{E}$ where are all Kraus operators are onsite and proportional to unitaries. Then co$(f)[\rho] \geq \textrm{co}(f)[\mathcal{E}[\rho]] $. In particular, under, such a channel, co$(QCMI)[\rho] \geq \textrm{co}(\textrm{QCMI})[\mathcal{E}[\rho]] $, i.e., co(QCMI) is non-increasing under the action of such a channel.}
     	\end{theorem}
     
     \begin{proof} 
     	 Let the optimal decomposition for $\rho$ vis-à-vis Def.~\ref{def:cof} be $\rho$ be $\rho = \sum_i p_i |\phi_i\rangle \langle \phi_i| $. This implies that $\mathcal{E}[\rho] = \sum_i \sum_\alpha q_\alpha U^{\dagger}_\alpha \left( \sum_i p_i |\phi_i\rangle \langle \phi_i| \right) U_\alpha$, where the Kraus operators are denoted as $\sqrt{q_\alpha} U_\alpha$ with $q_\alpha \geq 0$. Therefore, $\mathcal{E}[\rho] = \sum_{i \alpha} p_i q_\alpha |\tilde{\phi}_{i,\alpha}\rangle \langle \tilde{\phi}_{i,\alpha}|$, where $|\tilde{\phi}_{i,\alpha}\rangle = U_\alpha |\phi_i\rangle$. This expansion provides one decomposition for the density matrix $\mathcal{E}[\rho]$  which may or may not be optimal. By definition, co$(f)[\mathcal{E}[\rho] ] \leq \sum_{i,\alpha} p_i q_\alpha f(|\tilde{\phi}_{i,\alpha}\rangle) = \sum_{i,\alpha} p_i f(|\phi_i\rangle) = $ co$(f)[\rho]$ where we have used the fact that $f$ is invariant under onsite unitary transformations. Therefore, the convex-roof of any quantity $f$ that is invariant under onsite unitary transformations can't increase under such local decoherence. Since bipartite von Neumann entropy is invariant under onsite unitary transformations, and QCMI can be expressed as a linear combination of von Neumann entropies, this implies that co$(QCMI)[\rho] \geq \textrm{co}(QCMI)[\mathcal{E}[\rho]] $ under such a channel.
     	\end{proof}

     \begin{corollary}
     	Consider a density matrix  obtained by subjecting a pure state to onsite bit-flip or phase-flip decoherence with strength $p$. Let us denote the density matrix as $\rho(p)$ where $\rho(p=0)$ is the aforementioned pure state. If $p_2 > p_1$, then co(QCMI)$[\rho(p_1)] \geq $ co(QCMI)$[\rho(p_2)] $.
     	\end{corollary}
     	
     	\begin{proof}
     		\change{Bit-flip and phase-flip channels are closed under composition. Therefore, if $p_2 > p_1$, there exists a bit-flip/phase-flip channel $\mathcal{E}$ such that $\rho(p_2) = \mathcal{E}[\rho(p_1)]$. Since von Neumann entropy is invariant under onsite unitary transformations, and QCMI is just a linear combination of von Neumann entropies of different subregions, it is also invariant under onsite unitary transformations. Theorem~\ref{theorem:cof} then implies that co(QCMI)$[\rho(p_1)] \geq $ co(QCMI)$[\rho(p_2)] $.}
     	\end{proof}

   This rules out the scenario (b) in Fig.~\ref{fig:scenarios}.  In addition to co(QCMI), theorem \ref{theorem:cof} also applies to several other quantities of interest such R\'enyi versions of co(QCMI), or the convex-roof extension of bipartite R\'enyi/von Neumann entropies/mutual information. In strong contrast, the QCMI of the decohered density matrix is non-monotonic under local decoherence as explicitly demonstrated in Ref.~\cite{sang2024stability}. Perhaps more interestingly, this is also in contrast to pure ground state transitions out of topological states where the subleading universal part of von Neumann entanglement is generically larger at the quantum phase transition compared to the TEE of the proximate topological phase (as contrasted in Fig.~\ref{fig:CFTvsDEC}).

     
     \change{The most restrictive assumption in Theorem~\ref{theorem:cof} is the condition on Kraus operators being \textit{onsite} unitaries. Therefore, it is worth seeking potential generalizations. Clearly, there cannot be a vast generalization that allows for evolution under arbitrary quantum channels, since long-range entanglement can of course increase under long-depth unitary evolution (i.e. unitaries whose depth $d$ scales as $d \sim L$, where $L$ is the system's linear size). A natural generalization one might seek is evolution under Kraus operators that are all proportional to \textit{finite-depth} local unitary circuits, but are not restricted to be onsite (here finite-depth is a synonym for constant-depth, i.e., depth that does not scale with $L$.} If we assume that QCMI of a state is invariant under the action of a finite-depth local unitary circuit, then following the same argument as in the proof of Theorem~\ref{theorem:cof}, it follows that co(QCMI) is monotonically non-increasing if  Kraus operators that are all proportional to finite-depth local unitary circuits. However, there exist examples of `spurious TEE' \cite{bravyi2008spurious,cano2015interactions,liujun2016spurious}, which show that QCMI can change under the action of a finite-depth local unitary. \change{Although all known examples of spurious TEE are non-generic/fine-tuned, motivated from Ref.\cite{kim2023universal}, let us define a modified version of QCMI for a pure state $|\psi\rangle$:}
     
     \be 
     \change{\operatorname{QCMI}' (|\psi\rangle) = \textrm{inf}_U\left(\gamma(U|\psi_i\rangle)\right)}
     \ee  
    \change{where $\gamma(U|\psi_i\rangle)$ is the Levin-Wen QCMI of the state $U|\psi_i\rangle$ and infimum is taken over all possible finite-depth local unitaries $U$.  It is easy to see that  $\operatorname{QCMI}' (|\psi_1\rangle) = \operatorname{QCMI}' (|\psi_2\rangle)$ if $|\psi_1\rangle$ is related to $|\psi_2\rangle$ by a finite depth unitary.  It is then natural to define}
    
    \begin{equation}
    	\operatorname{co}( \operatorname{QCMI}' )[\rho] = \inf \left\{ \sum_i p_i  \operatorname{QCMI}'(|\psi\rangle) \right\}
    	\label{eq:def_coQCMIv2}
    \end{equation}
    where the  infimum as usual denotes minimization over all pure state decompositions of $\rho$ as  $\rho = \sum_i p_i |\psi_i\rangle \langle \psi_i|$ \cite{chaoming_discussion}. Based on this, we can now state the following proposition.

    \begin{theorem}\label{prop:coqcmi_monotonic}
    	 \change{Consider a density matrix that is subjected to a quantum channel $\mathcal{E}$ where are all Kraus operators are proportional to finite-depth local unitaries. Then co$(\operatorname{QCMI}')[\rho] \geq \textrm{co}(\operatorname{QCMI}')[\mathcal{E}[\rho]] $, i.e., $\operatorname{co}( \operatorname{QCMI}')$ is non-increasing under the action of such a channel.}
    \end{theorem}
      
           \begin{proof}
           	\change{The proof is essentially identical to the one for Theorem~\ref{theorem:cof}. Let's denote the action of the channel $\mathcal{E}$ as $\mathcal{E}[\rho] = \sum_\alpha q_\alpha U_\alpha \rho U^{\dagger}_\alpha$ where $U_\alpha$ are finite-depth local unitaries. If the optimal decomposition for $\operatorname{co}( \operatorname{QCMI}' )[\rho]$ is $\rho = \sum_i p_i |\psi_i\rangle \langle \psi_i|$, then under evolution by $\mathcal{E}$, $\rho \to \mathcal{E}[\rho] = \sum_{i,\alpha} p_i q_\alpha U_\alpha |\psi_i\rangle \langle \psi_i|U^{\dagger}_\alpha$. This provides one decomposition for $\mathcal{E}[\rho]$. Therefore, by definition of $\operatorname{co}( \operatorname{QCMI}')$, }
           	\bea 
          \operatorname{co}(\operatorname{QCMI}')[\mathcal{E}[\rho]] & \leq &  \sum_{i,\alpha} p_i q_\alpha \operatorname{QCMI}'(U_\alpha|\psi_i\rangle)\,\} \nonumber \\
           	& = & \sum_{i,\alpha} p_i q_\alpha \operatorname{QCMI}'(|\psi_i\rangle)\,\} \nonumber \\
           	& = & \sum_{i} p_i \operatorname{QCMI}'(|\psi_i\rangle)\,\} \nonumber \\
           	& = & 	\operatorname{co}(\operatorname{QCMI}')[\rho] 
           	\eea 
           \noindent where in the second equation we have used the invariance of $\operatorname{QCMI}'$ under a finite-depth local unitary. 
                      	\end{proof} 
           
           \change{We anticipate that in practice, $\operatorname{QCMI}'$ = QCMI for generic systems, since all known examples of spurious TEE are non-generic. This expectation is also supported by the calculation of QCMI in a large variety of lattice models and field theories \cite{Melko_Hubbard,jiang2012identifying,zhaoMeasuring2022,zhang2011topological,nielsen2012laughlin,laflorencie2016entanglement}. Therefore, we anticipate that the monotonicity result Theorem~\ref{prop:coqcmi_monotonic} will hold even for co(QCMI) and not just $\operatorname{co}(\operatorname{QCMI}')$.}
           
    As an aside, we note that in a gapped phase, the co(QCMI)/TEE takes its universal value only when the subsystem size $\ell$ that defines Levin-Wen partition satisfies $\ell \gg \xi$, where $\xi$ is the correlation length. However, the above monotonicity results hold true for any subsystem size irrespective of whether the condition $\ell \gg \xi$ is satisfied or not.

      \subsection{Can co(QCMI) be zero for $p < p_c$?}
      
     Zero co(QCMI) for the Levin-Wen partition would imply that the density matrix  admits a decomposition in terms of pure states with zero TEE, i.e., in Eq.~\eqref{eq:def_coQCMI} $\gamma(|\psi\rangle_i) = 0 \,\, \forall i$. Intuitively, one might think this is equivalent to the statement that the density matrix admits a decomposition in terms of SRE states. Indeed, zero TEE for 2d pure state implies that \textit{if} the pure state under consideration was the ground state of a gapped Hamiltonian, then the topological ground state degeneracy is zero \cite{kim2013longrange,kim2023universal}, but it need not imply that the state is SRE since the state may have long-range entanglement despite zero Levin-Wen QCMI. One such example is the GHZ state whose Levin-Wen QCMI vanishes, but the long-range entanglement captured via Kitaev-Preskill tripartite entropy \cite{kitaevTopological2006} does not vanish. Other possibility is that the state may have zero Levin-Wen QCMI, as well as zero Kitaev-Preskill tripartite entropy, but it may not be a ground state of a gapped Hamiltonian \cite{zhang2024unpublished}. Below, we will first focus on whether the density matrix can be written as a convex sum of  SRE pure states for $p< p_c$, and then consider whether co(QCMI) can be zero for $p < p_c$.

    As shown in Ref.~\cite{sang2023mixed,sang2024stability}, for $p < p_c$ there exists a constant time quasi-local Lindblad evolution $\mathcal{L}(\tau)$ that approximately converts the mixed state $\rho(p)$ to the pure toric code ground state $\rho(p=0)$. That is, 
     \be 
     \big|\mathcal{T} e^{\int_{0}^{1} \, dt\, \mathcal{L}(t)} \rho(p) - \rho(p=0)\big|_1 \leq \epsilon, \label{eq:tracenorm_maintext}
     \ee 
     where $\mathcal{T}$ denotes time-ordering, $|\cdot \big|_1$ denotes the trace norm and $\epsilon \ll 1$ is the tolerance (as discussed in Ref.~\cite{sang2024stability}, for a given $\epsilon$, the Lindblad evolution $\mathcal{L}$ corresponds to an $r$-local quantum channel where $r$ scales as $\log(\textrm{poly}(L)/\epsilon)$). 

     \begin{definition}\label{def:lowdepth} (Motivated from Ref.~\cite{sang2023mixed,sang2024stability}) A unitary circuit is `low-depth' if it consists of local gates state where the product of the maximal range of a gate times the depth of the circuit scales at most as polylog$(L)$ where $L$ is the system's linear size. Further, a pure state is short-range entangled (SRE) if it can be prepared via a low-depth local unitary circuit starting from a product state.
     \end{definition}
     
     \change{Before proceeding, we recall that a CSS (Calderbank-Shor-Steane) topological code \cite{css_Shor,css_Steane,kitaev2003fault,nielsen2002} is the ground state subspace of a topologically ordered system whose  Hamiltonian can be written in the form $H = H_X + H_Z$ where $H_X = \sum_i h_{i,X}$, and $H_Z = \sum_i h_{i,Z}$ satisfy the following properties. All local terms $\{h_{i,X}\}$ only involve Pauli-$X$ operators, and similarly all local terms $\{h_{i,Z}\}$ involve only Pauli-$Z$ operators. Further, all local terms $\{h_{i,X},h_{i,Z} \}$ mutually commute. In such a code, the topological degeneracy of $2^N$ on a torus arises from  pairs of logical operators $\{\overline{Z}_\alpha,\overline{X}_\alpha\}$ ($\alpha$ ranges from 1 to $N$ where $N$ depends on the code) with $\overline{Z}_\alpha \overline{X}_\alpha = - \overline{X}_\alpha \overline{Z}_\alpha$ (logical operators belonging to distinct $\alpha$ commute). A few examples of CSS topological codes are toric code \cite{kitaev2001unpaired} in any dimension, topological color codes \cite{bombin2006topological,bombin2007exact}, and the X-cube fracton model \cite{vijay2016fracton}.}
     
      \begin{theorem}\label{prop:longrange} \change{Consider any CSS topological code with geometrically local Hamiltonian in any spatial  dimension subjected to local decoherence at rate $p$. Let us denote the corresponding density matrix as $\rho(p)$. If $\rho(p)$ is a convex sum of SRE pure states, then a quasi-local Lindbladian  $\mathcal{L}$ that satisfies}
    \be 
     \big|\mathcal{T} e^{\int_{0}^{1} \, dt\, \mathcal{L}(t)} \rho(p) - \rho(p=0)\big|_1 \leq \tilde{\epsilon} \label{eq:tracenorm_x}
     \ee 
    \change{ must also satisfy}
     \be 
      \tilde{\epsilon} \geq (3-\sqrt{5})/2 \approx 0.38
     \ee.

    \end{theorem}

     \begin{figure}[t]
      \centering
      \includegraphics[width=0.9\columnwidth]{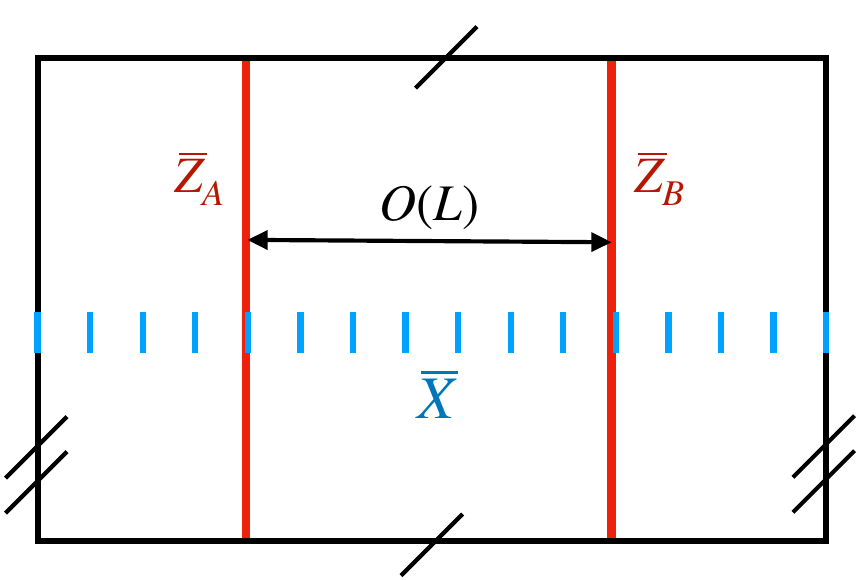}
      \caption{\change{Geometry used in the main text to show that the density matrix of a decohered CSS code can't be expressed as a convex sum of SRE pure states when $p < p_c$. Non-contractible logical operators $\overline{Z}_A$ and $\overline{Z}_B$ both anticommute with the non-contractible logical operator $\overline{X}$ (this figure is drawn for a 2d CSS code so that the logical operators are one-dimensional)}.}
      \label{fig:logical_strings}
      \end{figure}

     \begin{proof} We will sketch the main elements of the proof here, and refer the reader to Appendix \ref{sec:notSRE} for details. Let's assume that the condition (a) above is indeed satisfied. Therefore, we write
     \be 
     \rho(p) = \sum_a \, p_a |\xi_a\rangle \langle \xi_a|, \label{eq:SREdecompose}
     \ee 
     where $|\xi_a\rangle$ are SRE states.
     
     The main idea we employ is that the pure state $\rho(p=0)$ has long-range correlations of logical $Z$  operators (see, e.g., Ref.~\cite{jian2015long}), which we denote as $\overline{Z}_A$ and $\overline{Z}_B$ with the minimal distance between the operators $\overline{Z}_A,\overline{Z}_B$ $O(L)$ where $L$ is the total system's linear size. \change{In a non-trivial topological CSS code, there always exists a logical operator $\overline{X}$ that anticommutes and intersects with both $\overline{Z}_A,\overline{Z}_B$ (see Fig.~\ref{fig:logical_strings})}.   We chose $\rho(p=0)$ to be an eigenstate of $\overline{X}$.  The second result that we employ is that a low-depth local channel acting on a pure SRE state results in a density matrix whose connected correlations continue to be short-ranged \cite{lieb1972finite,poulin2010liebrobinson} for operators that are separated by distance of $O(L)$. That is, the connected correlations with respect to any of the mixed states $\rho_a = \mathcal{T} e^{\int_{0}^{1} \, dt\, \mathcal{L}(t)} \left(|\textrm{SRE}_a\rangle  \langle \textrm{SRE}_a|\right)$ must be short-ranged. Let's decompose the states $\rho_a$ into some set of pure states as $\rho_a = \sum_m q_{a,m} |\phi_{a,m}\rangle \langle \phi_{a,m}|$ (this decomposition is not unique, and it doesn't matter for our purposes what particular decomposition is chosen). Using Eq.~\eqref{eq:tracenorm_x}, we then arrive at the following set of equations:
     \bea
     \sum_{a,m} p_a q_{a,m} x_{a,m} \geq 1-\tilde{\epsilon} \nonumber \\
      \sum_{a,m,m'} p_a  q_{a,m}  q_{a,m'}z^A_{a,m} z^B_{a,m'} \geq 1-\tilde{\epsilon} \label{eq:implytracenorm_maintext}
    \eea 
      where $x_{a,m} =  \langle \phi_{a,m}|  \overline{X}| \phi_{a,m}\rangle$, $z^A_{a,m} =  \langle \phi_{a,m}|  \overline{Z}^A| \phi_{a,m}\rangle$ and $z^B_{a,m} =  \langle \phi_{a,m}|  \overline{Z}^B| \phi_{a,m}\rangle$. In addition, one has the following identity for any $a, m$: $\left(x_{a,m}\right)^2 + \left(z^A_{a,m}\right)^2 \leq 1$, and 
      $\left(x_{a,m}\right)^2 + \left(z^B_{a,m}\right)^2 \leq 1$ due to the anti-commutation relations between $\overline{X}$ and  $\overline{Z}_A, \overline{Z}_B$ \cite{toth2005entanglement}.
      As discussed in detail in Appendix \ref{sec:notSRE}, a series of Cauchy-Schwarz inequalities then imply that for aforementioned constraints to   hold simultaneously, the tolerance $\tilde{\epsilon}$ in Eq.~\eqref{eq:tracenorm_maintext} must satisfy $\tilde{\epsilon}  \geq (3-\sqrt{5})/2 \approx 0.38$.
      \end{proof}
      
      \begin{corollary} \label{corollary:longrange}
      \change{The density matrix of a CSS code subjected to local decoherence cannot be a convex sum of SRE pure states for $p < p_c$, where the threshold $p_c$ is defined via Eq.\ref{eq:tracenorm_maintext}.}
      \end{corollary}
      
      \begin{proof}
      	As shown in Ref.\cite{sang2024stability} the threshold $\epsilon$ in Eq.\ref{eq:tracenorm_maintext} can be made arbitrarily small as system size increases. In particular $\epsilon$ can be chosen to satisfy $\epsilon \sim 1/\textrm{poly}(L)$, while keeping the recovery channel $\mathcal{L}$ to be $\textrm{poly}(\log(L))$-local. Theorem~\ref{prop:longrange} then implies that $\rho$ cannot be a convex sum of SRE pure states for $p< p_c$.
      \end{proof}


     The above proof employed the fact that the pure state $\rho(p=0)$ has long-range correlations between the logical operators $\overline{Z}_A$, $\overline{Z}_B$  (Fig.\ref{fig:logical_strings}). Therefore, it also implies that $\rho(p)$ $p < p_c$ cannot be a convex sum of states where the two-point correlations of logical operators are short-ranged in each of the pure states $|\psi_i\rangle$ that enter the convex decomposition of $\rho(p)$. This shows that co(QCMI) must be non-zero for $p < p_c$ (thereby ruling out scenario (c)  in Fig.\ref{fig:scenarios}) since non-vanishing connected correlator of logical operators is a distinctive feature of topological order \cite{jian2015long}, and topologically ordered phases have non-zero Levin-Wen QCMI.

      \subsection{Does co(QCMI) equal $\log(2)$ for $p < p_c$?} Above we ruled out the possibility that the density matrix $\rho(p)$ can be written as a convex sum of SRE states for $p < p_c$. It is natural to ask how does the argument changes if one allows for a non-zero weight of topologically ordered state(s) in the convex decomposition of $\rho(p)$ for $p < p_c$. A natural possibility  is to consider  
     $\rho(p) = \sum_a \, p_a |\textrm{SRE}_a\rangle \langle \textrm{SRE}_a| + (1- \sum_a p_a) \rho(p=0)$, i.e., we allow a non-zero weight $1-\sum_a p_a \equiv 1 - w_{\text{SRE}}$ of topological ordered state in $\rho(p)$. As shown in Appendix \ref{sec:notSRE}, such a decomposition is not allowed if $\frac{\epsilon}{w_{\text{SRE}}} < (3-\sqrt{5})/2 \approx 0.38$ and therefore, as the tolerance $\epsilon \rightarrow 0$, the total weight of the SRE states in such a decomposition vanishes. This strongly suggests that co(QCMI) is not just non-zero for $p < p_c$, but equals $\log(2)$.  This would then rule out scenario (e) as well in Fig.~\ref{fig:scenarios}. Indeed, given any decomposition $\rho(p) = \sum_a p_a |\psi_a \rangle \langle \psi_a|$,  $\epsilon = 0$ implies that for all $a$, $\mathcal{T} e^{\int_{0}^{1} \, dt\, \mathcal{L}(t)} |\psi_a \rangle \langle \psi_a| = \rho(p=0)$, since $\rho(p=0)$ is a pure state. Following the same argument as above, then $|\psi_a\rangle$ can't be an SRE state since it is related to the toric code ground state via a low-depth local channel. This again strongly suggests that co(QCMI) equals log(2) for $p < p_c$.

    We now briefly discuss an alternative approach to co(QCMI) that exploits the average 1-form symmetry of the decohered toric code. This symmetry is generated by the operators $g_x = \prod_{e \in \ell} X_e$  that are product of Pauli operators $X_e$ along any closed-loop $\ell$ (including non-contractible ones). In Refs.~\cite{terhal2000entanglement, vollbrecht2001entanglement} Terhal, Vollbrecht and Werner developed an interesting scheme to calculate the convex-roof of any function $f$ of a density matrix $\rho$ that is symmetric under some group $G$, i.e., $g^{\dagger}_i \rho g_i = \rho$ where $g_i$ are the group elements of the group $G$. The basic idea is to exploit this symmetry to recast the problem of calculating the convex-roof into a different problem. First, one considers all possible pure states $|\psi_\alpha \rangle$, such that $\sum_i g^{\dagger}_i |\psi_\alpha \rangle \langle \psi_\alpha| g_i = \rho$ (one says that the pure state $|\psi_\alpha\rangle$ `twirls' to the mixed state $\rho$). Next, one minimizes the function $f$ over this set of pure states. Let us denote this minimum as $\epsilon(\rho)$. One can then show that the desired convex-roof co$(f)[\rho]$ equals the convex hull of $\epsilon(\rho)$, which is defined as the largest convex function on the set of all symmetric mixed states that nowhere exceeds $\epsilon$. This second step requires calculating the function $\epsilon(\rho_m)$ for all symmetric states $\rho_m$ in the neighborhood of the target state $\rho$. At least for a class of problems \cite{terhal2000entanglement, vollbrecht2001entanglement, manne2005entanglement}, the function $\epsilon(\rho)$ is already convex, so this second step is redundant, and one only needs to minimize $f$ over pure states that twirl to  $\rho$. 
     
     Following this idea and exploiting the average 1-form symmetry of the decohered toric code density matrix, the first step is to minimize  TEE over all \textit{pure} states that twirl to $\rho$, i.e., minimize TEE over all states $|\phi\rangle$ such that $\sum_x g_x |\phi \rangle \langle \phi| g_x = \rho$ where $g_x$ are the aforementioned operators that form closed loops. Following the same argument as above, any such pure state $|\phi\rangle$ cannot be SRE for $p < p_c$ since $g_x$ can be thought of as Kraus operators corresponding to a finite-depth channel. Furthermore, again following the same argument as above, $|\phi\rangle$ should inherit the long-range order associated with non-contractible logical string operators from $\rho$. Therefore, it is reasonable to expect that the minimal TEE for any such state $|\phi\rangle$ is the one corresponding to the pure toric code, i.e., $\log(2)$. Remarkably, the decomposition of $\rho$ discussed in Ref.~\cite{Chen_Separability_2024} that captures the separability transition already takes the desired form, namely, $\rho(p) = \sum_{g_x} |\psi_{g_x}(p)\rangle \langle \psi_{g_x}(p)|$
where  $|\psi_{g_x}(p)\rangle = g_x |\psi(p)\rangle$, where the pure state $|\psi(p)\rangle$ was argued to have TEE of $\log(2)$ for $p < p_c$ (we will discuss the pure state $|\psi(p)\rangle$ in detail in Sec.~\ref{sec:numerics} and verify that it is indeed $\mathbb{Z}_2$ topologically ordered for $p < p_c$). Since the mixed state is topologically ordered for $p < p_c$, and topological order is expected to be robust to small perturbations, it is reasonable to expect that the minimum TEE of a pure state that twirls to any symmetric mixed state in the neighborhood of $\rho$ is also $\log(2)$. This again suggests that co(QCMI) of $\rho$ equals $\log(2)$ for $p < p_c$. Admittedly, this argument is heuristic, and it will be worthwhile to pursue it further.

     \subsection{Is co(QCMI) zero for $p > p_c$?} Ref.~\cite{Chen_Separability_2024} provided an explicit decomposition of the decohered toric code density matrix that was argued to consist of states with zero TEE for $p> p_c$. Since the statistical mechanics model involved in the corresponding argument is not exactly solvable to our knowledge, it is desirable to supplement the analytical treatment of Ref.~\cite{Chen_Separability_2024} with a direct numerical simulation. In Sec.\ref{sec:numerics}, we will perform a tensor-assisted Monte Carlo (TMC) simulation to extract the R\'enyi co(QCMI) for the decomposition in Ref.~\cite{Chen_Separability_2024}, and provide direct evidence that R\'enyi co(QCMI) indeed vanishes for $p > p_c$.

     Combining all the arguments in this section, we therefore conclude that co(QCMI) is non-zero for $p < p_c$, and zero for $p > p_c$. Further, we expect that in fact it equals $\log(2)$ for $p < p_c$ (scenario (a) in Fig.~\ref{fig:scenarios}). We now turn to a specific decomposition of the density matrix that we will conjecture is optimal for co(QCMI), i.e., it achieves the minimum in Eq.~\eqref{eq:def_coQCMI}. As we will see, this conjecture is equivalent to the statement that the co(QCMI) of the decohered toric code equals the TEE of a specific pure state $\ket{\psi}$. To this end, we develop a tensor-assisted Monte Carlo (TMC) computational scheme to efficiently evaluate the R\'enyi TEE of the aforementioned pure state. The TMC method is designed to significantly mitigate the exponential complexity associated with numerical evaluation of R\'enyi TEE ~\cite{zhangIntegral2024,zhouIncremental2024}, and allows us to study the behavior of R\'enyi TEE close to the transition.
     
\section{Testing the conjectured `optimal decomposition' for co(QCMI) using many-body simulations} \label{sec:numerics}

\change{
In this section, we adopt the convention in Ref.~\cite{Chen_Separability_2024}, where bold font $\mathbf{e}$ in $x_\mathbf{e}$ denotes the collection of bond on all edges, while $x_e$ denotes the bond on a specific edge $e$.
}

\subsection{Conjectured optimal decomposition} \label{sec:conjecture}

Ref.~\cite{Chen_Separability_2024} proposed a specific decomposition of the decohered 2d toric code density matrix $\rho(p)$, arguing it correctly captures the location and the universality class of the transition. Concretely, this decomposition is given as \be 
\rho(p) = \sum_{g_x} |\psi_{g_x}(\beta)\rangle \langle \psi_{g_x}(\beta)| \label{eq:optimaldecompose},
\ee
where  $|\psi_{g_x}(\beta)\rangle = g_x |\psi(\beta)\rangle$,  $g_x$  is a product of single-site Pauli-$X$ operators that form closed loops, and the (unnormalized) pure state $|\psi(\beta)\rangle$ is given by 
\be 
\ket{\psi(\beta)}\propto\sum_{x_\mathbf{e}}\sqrt{\pf{x_\mathbf{e}}(\beta)}\ket{x_\mathbf{e}}, \label{eq:optimalpurestate}
\ee 
where $\pf{x_\mathbf{e}}(\beta)=\sum_{z_{\textbf{v}}}e^{\beta \sum_e x_e\prod_{v\in e}z_v}$ is the partition function of the 2d Ising model with bond strengths given by Ising variables $x_{\textbf{e}}$. The relation between the inverse temperature $\beta$ and the decoherence rate $p$ is $\tanh(\beta) = 1- 2p$. Notably, this wavefunction precisely corresponds to the toric code ground state at $\beta = \infty$ and to the product state in the $Z$-basis at $\beta = 0$. 

Since states $|\psi_{g_x}(\beta)\rangle$ are all related to the state $\ket{\psi(\beta)}$ via onsite-unitaries, the decomposition in Eq.~\eqref{eq:optimaldecompose} implies that $\textrm{co(QCMI)}[\rho(p)] \leq \textrm{TEE}(\ket{\psi(\beta)})$. We conjecture that the decomposition in Eq.~\eqref{eq:optimaldecompose} is optimal for co(QCMI), i.e., it achieves the minimum of average TEE over all possible decompositions of the density matrix (see Eq.~\eqref{eq:def_coQCMI}).  This conjecture is equivalent to the statement that 
\be 
\textrm{co(QCMI)}[\rho(p)] \stackrel{?}{=} \textrm{TEE}(\ket{\psi(\beta)}) \label{eq:conjecture}
\ee 

In this section, we will test this conjecture by subjecting the right hand side of the above equation to the constraints derived in the last section. The main motivation for our conjecture comes from Ref.~\cite{Chen_Separability_2024} which argued that the TEE for the state $\ket{\psi(\beta)}$ jumps from $\log(2)$ to zero at $p = p_c$, in line with our expectation for the co(QCMI) of $\rho(p)$ (Sec.~\ref{sec:coQCMIproperties}). However, the results in Ref.~\cite{Chen_Separability_2024} were not strong enough to test whether the R\'enyi TEE of $\ket{\psi(\beta)}$ is monotonic as a function of $p$, especially in the vicinity of $p_c$. Therefore, we will develop unbiased quantum many-body numerical techniques to calculate the R\'enyi TEE of $\ket{\psi(\beta)}$.

\subsection{Anyon condensation order parameter} \label{sec:anyoncondensation}
 Before we discuss the R\'enyi TEE of the state $\ket{\psi(\beta)}$, we compute and verify the universal properties of the decoherence induced transition in noisy toric code from the \textit{single} pure state  $\ket{\psi(\beta)}$.
 
\begin{figure}[htp!]
\centering
\includegraphics[width=\columnwidth]{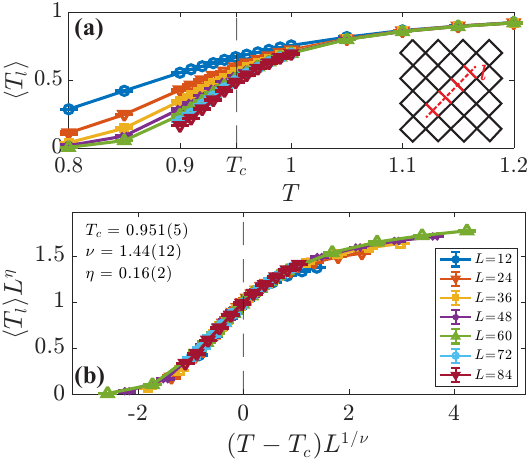}
\caption{\textbf{Anyon condensation operator.} (a) Average value of the anyon condensation order parameter $\langle T_l\rangle$ with respect to the  state $|\psi(\beta)\rangle$ (Eq.\ref{eq:optimalpurestate}) as a function of the temperature $T \,(=\beta^{-1})$. The operator $T_l$ is depicted by the red dashed segment of length $l$. (b) Data collapse of $\langle T_l\rangle$. The two panels share the same legend as the one shown in (b).}
\label{fig:result_AC}
\end{figure}

 The universal aspects of the decoherence induced transition in toric code are known to be related to the Nishimori multicritical point \cite{nishimori1981internal,dennis2002,wang2003confinement,fan2023diagnostics,lee2023quantum}. One calculation that is suggested by the results in Ref.~\cite{Chen_Separability_2024} is that of the anyon condensation order parameter with respect to  the state $\ket{\psi(\beta)}$. Concretely, one considers a path $l$, and calculates the expectation value of the operator $T_l=\prod_{e\in l}Z_e$ that flips the bonds on path $l$ (when the wavefunction is expressed in the $X$ basis, as in Eq.~\eqref{eq:optimalpurestate}). This expectation value captures the tunneling amplitude between different logical sectors of the toric code. In the topological phase (i.e. $p < p_c$), one expects that this tunneling amplitude vanishes, while in the non-topological phase (i.e. $p > p_c$), one expects that it will be non-zero \cite{Chen_Separability_2024}. If the wavefunction $\ket{\psi(\beta)}$ correctly captures the universal aspects of Nishimori multicritical  point, then using Wegner duality \cite{wegner1971duality} one expects the following behavior in the vicinity of the critical point:
 \be 
\langle \psi(\beta) |T_l|\psi(\beta)\rangle \sim f(L/\xi)/L^{\eta} \label{eq:Tscaling}
 \ee 
\change{where $\eta$ will turn out to be the anomalous dimension associated with a specific moment of the disorder operator correlator at the random bond Ising model's (RBIM) Nishimori multicritical point (as discussed below)}, $\xi \sim (p-p_c)^{-\nu}$ is the corresponding diverging correlation length, and we have chosen the length of the path $l$ to be proportional to the total system's linear size $L$ (as depicted in the inset of Fig.~\ref{fig:result_AC} (a)). It is worth noting that the expectation value of $T_l$ with respect to the decohered density matrix $\rho(p)$ itself, i.e., $\tr(\rho(p)T_l) =\sum_{g_x} \langle \psi_{g_x}(\beta)|T_l|\psi_{g_x}(\beta)\rangle$,  will \textit{not} see any singularity across the transition. This is because the transition is not visible in any quantity linear in the density matrix \cite{fan2023diagnostics}. The object of our study, namely $\langle \psi(\beta) |T_l|\psi(\beta)\rangle$, equals (upto a sign) $\sum_{g_x} |\langle \psi_{g_x}(\beta)|T_l|\psi_{g_x}(\beta)\rangle|$, and cannot be expressed as a linear function of the density matrix.

Next, we discuss the details of the numerical evaluation of $\langle \psi(\beta) |T_l|\psi(\beta)\rangle \equiv \langle T_l \rangle$. We consider a system with open boundaries and chose the path $l$ as a line segment on the dual lattice, as shown by the red dashed line in Fig~\ref{fig:result_AC} (a). Its expectation value can then be computed as \cite{Chen_Separability_2024}
\be
\begin{split}
T_l\ket{\psi}\propto\sum_{x_\mathbf{e}}&\sqrt{\pf{x_\mathbf{e}}}\ket{x_{\textbf{e}_l}} \text{, and}\\
\av{T_l}=\frac{\sum_{x_\mathbf{e}}\sqrt{\pf{x_\mathbf{e}}\pf{x_{\textbf{e}_l}}}}{\sum_{x_\mathbf{e}}\pf{x_\mathbf{e}}}
&=\frac{\sum_{x_\mathbf{e}}\pf{x_\mathbf{e}}\sqrt{\frac{\pf{x_{\textbf{e}_l}}}{\pf{x_\mathbf{e}}}}}{\sum_{x_\mathbf{e}}\pf{x_\mathbf{e}}},\\
&=\left [\sqrt{\frac{\pf{x_{\textbf{e}_l}}}{\pf{x_\mathbf{e}}}}\;\right ]
\end{split}
\ee
where the extra subscript in $x_{\textbf{e}_l}$ represents the flipped bond configuration on line segment $l$, and $[\,\cdot\,]$ denotes the weighted average over bond configurations with probability proportional to the corresponding partition function $\pf{x_\mathbf{e}}$, i.e., $W(x_\mathbf{e})\propto\pf{x_\mathbf{e}}(\beta)=\sum_{z_{\textbf{v}}}e^{\beta \sum_e x_e\prod_{v\in e}z_v}$.  \change{$\av{T_l}$ precisely corresponds to the two-point correlator $[\langle \mu(\vec{0}) \mu(\vec{r})\rangle^{1/2}] $ of the disorder operator $\mu$ in the RBIM studied in Ref.\cite{merz2002negative}, where $\langle\,\cdot\,\rangle$ denotes the average with respect to a fixed disorder configuration, and $[\,\cdot\,]$ again denotes disorder averaging. $\vec{0}$ and $\vec{r}$ are the two end-points of $T_l$. The generation of bonds according to the probability distribution $\pf{x_\mathbf{e}}$ can be greatly simplified using the Nishimori condition $\tanh(\beta)=1-2p$. In particular, one simply proposes an updated bond configuration according to the binomial distribution with probability $p$ at each MC step, followed by the gauge transformation ${x_e\;\rightarrow\;x_e\prod_{v\in e}\sigma_v}$ with ${\sigma_v=\pm 1}$ on every site. By doing so, one generates a bond configuration according to the distribution $W'(x_\mathbf{e})\propto \textstyle\sum _{{[x_\mathbf{e}]}}  {\textstyle \prod_{e}p^{\delta_{AFM}(x_e)}(1-p)^{\delta_{FM}(x_e)}} $, where  $[x_\mathbf{e}]$ denotes an equivalence class of $2^N$ bond configurations related to $x_\mathbf{e}$ by aforementioned gauge transformations, and $\delta_{AFM(FM)}(x_e)=1$  if bond $x_e$ is AFM(FM), otherwise $0$. Summing up all $2^N$ terms related to each other via gauge transformation gives $W'(x_\mathbf{e})=\sum_{\sigma_\mathbf{v}} \prod_{e}p^{\delta_\textrm{AFM}(x_e\prod_{v\in e}\sigma_v)}(1-p)^{\delta_\textrm{FM}(x_e\prod_{v\in e}\sigma_v)}\propto\sum_{\sigma_\mathbf{v}}e^{\beta \sum_e x_e\prod_{v\in e}\sigma_v}=\pf{x_\mathbf{e}}$, where we have used the  Nishimori relation $\tanh(\beta)=1-2p$.}



\begin{figure}[htp!]
\centering
\includegraphics[width=\columnwidth]{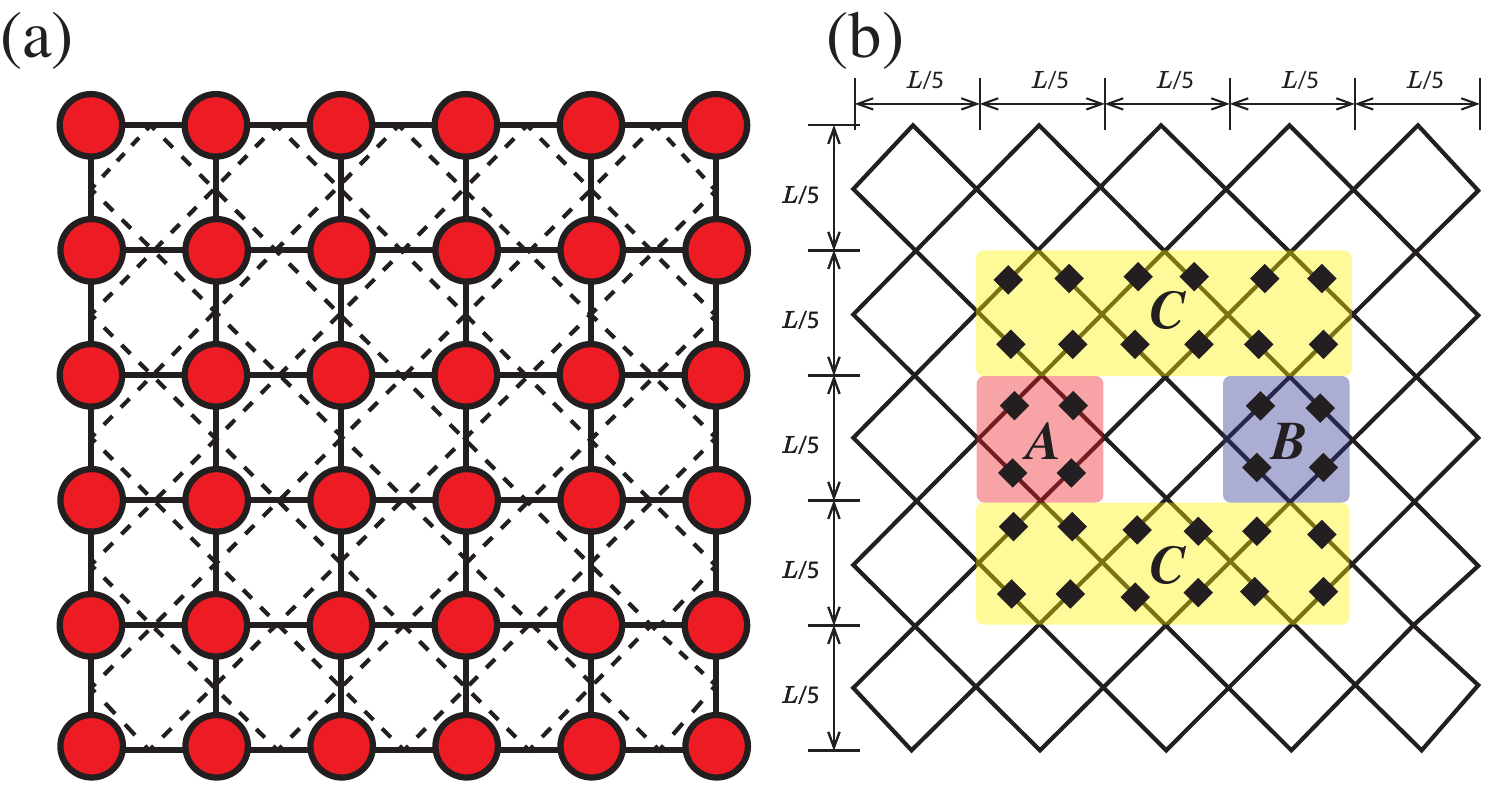}
\caption{\textbf{Tensor formalism for the Ising model partition function.} (a) A tensor network for $5\times 5$ system, with $(L+1)^2=36$ local tensors. Each tensor encodes the interaction between 4 Ising spins, with each leg containing the local spin degree of freedom ($d=2$). The dashed lines represent the original lattice. (b) The Levin-Wen scheme in the tilted square lattice. Here $L$ is a multiple of 5, and we divide the system into a $5\times5$ grid, and choose the subregions as depicted, similar to Ref.~\cite{Wu_Entanglement_2020}.}
\label{fig:tensor}
\end{figure}

To calculate the partition function $\pf{x_\mathbf{e}}$ for a given bond configuration $x_\mathbf{e}$, we performed the matrix product state (MPS) assisted sampling with bond dimension $\chi=8$ (we will provide details of this method in Sec.~\ref{sec:TMC}). Such a small bond dimension is sufficient to capture the Ising partition function, with relative error of order $10^{-20}$. For simplicity in the MPS calculation, we chose our system to be the tilted square lattice with open boundary condition (OBC), as shown in Fig.~\ref{fig:tensor} and the inset of Fig.~\ref{fig:result_AC} (a)).

Fig.~\ref{fig:result_AC} (a) shows the result of the anyon condensation operator. It approaches zero in the topological phase and gradually increases as a function of $T = \beta^{-1}$. Performing finite size scaling following Eq.~\eqref{eq:Tscaling} leads to a rather accurate data collapse with $T_c=0.951(5)$, $\nu=1.44(12)$ and $\eta=0.16(2)$, see Appendix~\ref{sec:AC_CE} for details of the data collapse procedure.  This set of numbers agrees well with the previous studies on RBIM along the Nishimori line (\change{see TABLE.~\ref{table:exponents}}) suggesting that the conjectured optimal decomposition  of the decohered density matrix (Eq.~\eqref{eq:optimaldecompose}) indeed correctly captures the decoherence induced transition. With such verification at hand, we now turn to the discussion of the TEE for the state $|\psi(\beta)\rangle$.

\begin{table}[h!]
\centering
\begin{tabular}{|c|c|c|c|}
\hline
%
%
$T_c$	&$\nu$		&$\eta$		&Reference						\\ \hline
0.9538(9) 	& 1.33(3)		& 	-	&\cite{honecker2001universality}	 	\\ \hline
0.9533(9) 	& 1.50(3)		& 	-	&\cite{merz2002twod} 				\\ \hline
0.9533(9) 	& 1.48(3)		&  -	&\cite{picco2006strong} 				\\ \hline
0.9528(4) 	& 1.52(3)		& 	-	&\cite{Hasenbusch_Multicritical_2008} 	\\ \hline
-  	&  	-	& 0.17(3)		&\cite{merz2002negative} 	\\ \hline
0.951(5) 	& 1.44(12) 	&0.16(2)		&This work 						\\ \hline
\end{tabular}
\caption{\textbf{\change{Critical exponents for the random bond Ising model at the Nishimori critical point}.}}
\label{table:exponents}
\end{table}

\subsection{Tensor-assisted Monte Carlo method for R\'enyi entanglement entropy} \label{sec:TMC}

As discussed above, our aim is to numerically calculate the TEE of the state $|\psi(\beta)\rangle$ to test Eq.~\eqref{eq:conjecture}. Numerically, calculating the von Neumann entanglement $-\tr (\rho \log \rho)$ for a density matrix $\rho$ is rather challenging for generic 2D systems. Instead, we will focus on the numerical evaluation of the   R\'enyi TEE, which can be argued to equal von Neumann TEE for topologically ordered pure states, using field-theoretic and lattice-based arguments \cite{dong2008topological,flammia2009topological,zhang2012quasiparticle}. Perhaps more pertinently, the  R\'enyi version of the co(QCMI) also satisfies the monotonicity discussed in Sec.~\ref{sec:coQCMIproperties}, and therefore, provides as good a test for monotonicity as the von Neumann based co(QCMI). 

Since we develop a new technique, namely tensor-assisted Monte Carlo (TMC) method for calculating the R\'enyi TEE, in this subsection we will describe the method in detail, and discuss the results in the next subsection.

The second order R\'enyi entropy for a density matrix $\rho_A$ is defined as $S_2=-\ln\tr{\rho_A^2}$.  Defining $S_2$ requires dividing the total system into subregions $A$ and $B$. \change{Let us label the bond configuration $x_\mathbf{e}$ for the whole system as $x_\mathbf{e} \equiv (x_A, x_B)$, where $x_A, x_B$ denote the configurations in subregion $A$ and $B$ respectively. } For the wavefunction in Eq.~\eqref{eq:optimalpurestate} one finds,

\begin{eqnarray}
\tr{\rho_A^2} & = & \frac{\sum_{x_{\textbf{e}}x'_{\textbf{e}}}\left(\pf{x_A,x_B}\pf{x'_A,x'_B} \pf{x'_A,x_B}\pf{x_A,x'_B}\right)^{1/2}}{\sum_{x_{\textbf{e}}x'_{\textbf{e}}}\pf{x_A,x_B}\pf{x'_A,x'_B}}\nonumber \\
 &  = &\left [\sqrt{\frac{\pf{x'_A,x_B}\pf{x_A,x'_B}}{\pf{x_A,x_B}\pf{x'_A,x'_B}}}\;\right ]
\label{eq:rho2}
\end{eqnarray}
\change{where the sum $\sum_{x_{\textbf{e}}x'_{\textbf{e}}}$ runs over two replicas of the system, and $[\,\cdot\,]$  denotes the weighted average over bond configurations with the joint probability proportional to $\pf{x_\mathbf{e}} \pf{x'_\mathbf{e}} $. Note that $\pf{x_\mathbf{e}} \equiv \pf{x_A,x_B}$ and $\pf{x'_\mathbf{e}} \equiv \pf{x'_A,x'_B}$, while $\pf{x'_A,x_B} \pf{x_A,x'_B}$ denotes the partition function of the two copies where the bond configuration in region A has been swapped between the two copies.}

Our goal is to perform the averaging in Eq.~\eqref{eq:rho2} using Monte Carlo sampling. At low temperatures, the direct sampling becomes exponentially hard. This is because the quantity being sampled in Eq.~\eqref{eq:rho2} is around 1 for only an exponentially small set of the total configurations, and approximately zero otherwise. This is the typical problem of sampling an exponential observable that plagues other EE computations~\cite{panStable2023,liaoControllable2023,zhangIntegral2024,zhouIncremental2024}, and is related to the log-normal distribution of the quantity being sampled. If EE for a subregion $A$ follows the area-law ($S_A \propto L_A$ in 2d), naive Monte Carlo sampling results in an exponentially increasing relative error in measuring $ e^{-S_2} $, rendering any quantitative estimate of EE extremely difficult. 
To overcome such problem of exponential observables, many incremental methods, with the hope to mitigate the exponentially computational complexity~\cite{kallinAnomalies2011,drutHybrid2015,drutEntanglement2016,albaOut2017,D'Emidio_Entanglement_2020,zhaoMeasuring2022,zhaoScaling2022,songExtracting2023}, have been put forward over the years. It has been argued that at least for the 2d Hubbard and the Heisenberg models, there is a systematic procedure to convert the exponential complexity to a polynomial one as discussed in ~\cite{liaoControllable2023,zhangIntegral2024,zhouIncremental2024}. In the TMC method, we adopt  a  similar strategy and combine it with a tensor-network approach to speed up calculations. In particular, for each bond configuration $\{x_\mathbf{e},x'_\mathbf{e}\}$, the corresponding 2d Ising partition functions can be quickly evaluated via the contraction of a 2d tensor-network. We can then sample the bond configurations $\{x_\mathbf{e},x'_\mathbf{e}\}$ using standard Monte Carlo procedure. The TMC method hence combines the contraction of the MPS for any fixed random bond configuration and the non-equilibrium Monte Carlo sampling~\cite{albaOut2017,D'Emidio_Entanglement_2020,zhaoMeasuring2022} of the random bond configurations.

The details of the algorithm are as follows. First, consider the following object $Q(\lambda)$:
\begin{equation}
\begin{split}
Q(\lambda)&=\sum_{x_{\textbf{e}}x'_{\textbf{e}}}\pf{x_A,x_B}\pf{x'_A,x'_B}\left (\frac{\pf{x'_A,x_B}\pf{x_A,x'_B}}{\pf{x_A,x_B}\pf{x'_A,x'_B}}\right )^{\lambda/2}\\
&=\sum_{x_{\textbf{e}}x'_{\textbf{e}}}\pf{x_A,x_B}\pf{x'_A,x'_B} g\left(x_\mathbf{e},x'_\mathbf{e},\lambda\right),
\label{eq:TMC_pf}
\end{split}
\end{equation}
\change{where $g\left(x_\mathbf{e},x'_\mathbf{e},\lambda\right)=\left (\frac{\pf{x'_A,x_B}\pf{x_A,x'_B}}{\pf{x_A,x_B}\pf{x'_A,x'_B}}\right )^{\lambda/2}$. The second Renyi entropy (Eq.~\eqref{eq:rho2}) is then given by $e^{-S_2} = Q(1)/Q(0)$.} The advantage of this formulation is that we can now use the Jarzynski equality~\cite{Jarzynski_Nonequilibrium_1997,albaOut2017,D'Emidio_Entanglement_2020}, namely, the exponential of the free energy difference equals the weighted average of the exponential of the work done over all realizations bringing the system from $Q(0)$ to $Q(1)$:

\begin{equation}
\begin{split}
e^{-S_2}&=\frac{Q(1)}{Q(0)}=e^{\Delta F}=\av{e^W} \text{, with}\\
W&=\int_0^1\mathrm{d}\lambda\pdiff{\ln g\left(x_\mathbf{e},x'_\mathbf{e},\lambda\right)}{\lambda}\\
\label{eq:TMC_work}
\end{split}
\end{equation}
Since all partition functions $\pf{}$ can be computed  by contracting corresponding tensor network, using the  updating scheme from Sec.~\ref{sec:anyoncondensation}, one can propose configurations from the joint probability $\pf{x_A,x_B}\pf{x'_A,x'_B}$. By additionally choosing acceptance probability as ratio of $g\left(x_\mathbf{e},x'_\mathbf{e},\lambda\right)$ (Eq.~\eqref{eq:TMC_pf}) between the new and old bond configurations, one can sample under the distribution $Q(\lambda)$. 

For each TMC calculation, one gradually increases $\lambda$ from 0 to 1, updates bond configurations $\{x_\mathbf{e},x'_\mathbf{e}\}$ according to $Q(\lambda)$ and measures the infinitesimal work done $\change{\mathrm{d}W=\ln\sqrt{\frac{\pf{x'_A,x_B}\pf{x_A,x'_B}}{\pf{x_A,x_B}\pf{x'_A,x'_B}}}\mathrm{d}\lambda}$ accumulated in each step (Eq.~\eqref{eq:TMC_work}). The final EE is simply $S_2=-\ln(\av{e^W})$. Previous works argued that scaling the number of discretization steps with system size leads to a polynomial time algorithm for calculating EE while keeping the relative error fixed ~\cite{liaoControllable2023,zhouIncremental2024,zhangIntegral2024}. In this work however, we  use a fixed but sufficiently large number of discretization steps ($= 2\times10^5$), independent of the system size. This choice gives a satisfactory relative error for TEE close to the critical point for the system sizes studied, as we now discuss.

\subsection{Results for R\'enyi TEE} \label{subsec:result_TEE}
We now discuss the numerical results for the R\'enyi TEE of the state $|\psi(\beta)\rangle$ obtained using the aforementioned TMC method. The Levin-Wen partition to define TEE is shown in Fig.~\ref{fig:tensor} with  R\'enyi TEE given by $\gamma = \frac{1}{2}S_2(A:B|C) = \frac{1}{2}\left( S_2(AC) + S_2(BC) - S_2(C) - S_2(ABC)\right) $. To perform finite-size scaling, we maintain the shapes of the regions A, B, C and scale the total system size so that each of the regions $A, B, C$ scale with $L$. We simulated the linear system sizes $L=5,10,15$ and the temperature $T\in [0.2,1.2]$ with data points that lie on the either side of the critical point $T_c=0.954(6)$ which is determined from the anyon condensation operator in Fig.~\ref{fig:result_AC}. 

\begin{figure}[htp!]
\centering
\includegraphics[width=\columnwidth]{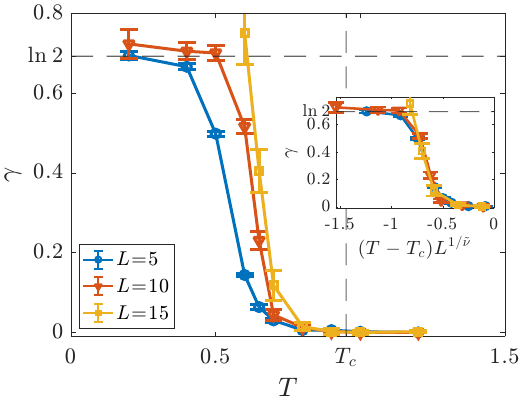}
\caption{\textbf{Result for R\'enyi TEE $\gamma$ using Levin-Wen scheme.} R\'enyi TEE $\gamma$ for the state  $|\psi(\beta)\rangle$ (Eq.\ref{eq:optimalpurestate}) against temperature $T \,(=\beta^{-1})$ and the rescaled temperature $(T-T_c)L^{1/\tilde{\nu}}$ (inset) with $T_c=0.951$ and $\tilde{\nu}\approx3.2$.}
\label{fig:result_gamma}
\end{figure}

Fig.~\ref{fig:result_gamma} shows the numerically obtained R\'enyi TEE $\gamma$. Again recall that the temperature $T$ is related to the decoherence rate $p$ via $\tanh(1/T) = 1- 2p$. The overall trends are as follows. $\gamma \approx \ln 2$ at low temperatures for all system sizes, $\gamma$ is monotonically non-increasing as $p$ increases, and it tends towards zero as $T \to T_c$. Further, as the system size is increased, $\gamma$ tends towards $\log(2)$ at a relatively higher temperature and is also non-zero up till a relatively higher temperature (i.e., the range of decoherence rate over which the topological phase is visible in a finite system increases). These numerical results rule out scenarios (b) and (d) in Fig.~\ref{fig:scenarios} for the R\'enyi co(QCMI), which is consistent with the analytical arguments in Sec.~\ref{sec:coQCMIproperties} and Ref.~\cite{Chen_Separability_2024}. Perhaps more interestingly, they strongly suggest that as one approaches the critical point, so that $L \ll \xi$, R\'enyi co(QCMI) approaches zero. Assuming that the von Neumann TEE has the same qualitative behavior as the R\'enyi TEE \cite{dong2008topological,flammia2009topological,zhang2012quasiparticle}, this indicates that the von Neumann co(QCMI) also approaches zero as $p \rightarrow p_c$ (recall that the TEE of the state $|\psi(\beta)\rangle$ puts an upper bound on the co(QCMI), and the von Neumann TEE is necessarily non-negative due to strong subadditivity). This is in strong contrast to (pure) ground state phase transition in toric code that is driven by a magnetic field, where in the critical regime, QCMI \textit{exceeds} the TEE of the topological phase. See Fig.~\ref{fig:CFTvsDEC} for a contrast between the pure state transition and the decoherence induced transition. 
Overall, our results are consistent with the scenario (a) in Fig.~\ref{fig:scenarios} in the thermodynamic limit, in line with the analytical arguments in Sec.~\ref{sec:coQCMIproperties} and our conjecture relating TEE of $|\psi(\beta)\rangle$ to the co(QCMI) of the decohered state (Eq.~\eqref{eq:conjecture}).

We also attempted finite-size scaling for TEE with the  scaling form $\gamma(T,L)=f\left ((T-T_c)L^{1/\tilde{\nu}}\right )$. We found that $\gamma$ obtained for different system sizes collapses well when we choose $\tilde{\nu}\approx 3.2$, see the inset of Fig.~\ref{fig:result_gamma}. This value is much larger than the critical exponent $\nu$ for the Nishimori critical point (namely $\nu \approx 1.5$, which agrees well with the exponent obtained from anyon condensation order parameter from the same wavefunction, as discussed in Sec.~\ref{sec:anyoncondensation}). We suspect this discrepancy is partly because the system sizes for which we can access TEE is still limited, and perhaps the finite-size effects for TEE are also relatively larger compared to those for the anyon condensation order parameter. Furthermore, in the critical regime $(L/\xi \ll 1)$, we only have a few data points. Nonetheless, the data collapse is suggestive that TEE is a function only of $L/\xi$, where $\xi$ is the diverging correlation length. Although we don't have an analytical understanding of TEE close to the transition, arguments in  Ref.~\cite{Chen_Separability_2024} imply that the TEE is related to the domain-wall free energy in the RBIM along the Nishimori line, which scales as $(L/\xi)^{1/\nu}$ close to the transition \cite{honecker2001universality}. This motivates a scaling ansatz in the critical regime ($L \ll \xi $) of the form $\gamma(L/\xi) = \log(2)\left(1 - \frac{\log(1 + a e^{-b(L/\xi)^{1/\nu}})}{\log(1+a)}\right)$ where $a, b$ are some numbers. Such a scaling form is also suggested from previous works on topological entanglement negativity in thermal or decoherence driven topological transitions \cite{lu2023characterizing,lu2024disentangling}. Taylor expanding such an expression would then imply $\gamma \sim (L/\xi)^{1/\nu} \sim L^{1/\nu} (T_c-T)$, which at least qualitatively captures the features of our numerical results. For example, at a fixed $p \lesssim p_c$, $\gamma$ is an increasing function of $L$.

\begin{figure}[htp!]
\centering
\includegraphics[width=\columnwidth]{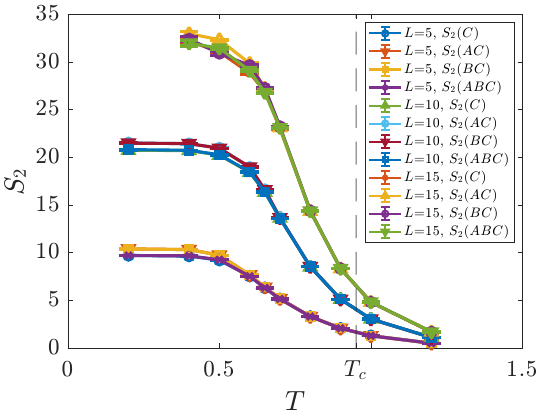}
\caption{\textbf{Second order R\'enyi entanglement entropy $S_2$.} R\'enyi entanglement entropy $S_2$ for the state $|\psi(\beta)\rangle$ (Eq.\ref{eq:optimalpurestate}) against temperature $T$ for the four subregions depicted in Fig.~\ref{fig:tensor}. $S_2$ for all of these regions is 
monotonically non-increasing as a function of increasing temperature/decoherence rate.}
\label{fig:result_S}
\end{figure}

Let us also briefly discuss the behavior of the bipartite  R\'enyi entropy in the pure state $|\psi(\beta)\rangle$. As shown in Fig.~\ref{fig:result_S}, for all subregions we looked at, $S_2$ is a monotonically decreasing function of the decoherence rate. Recall that the results in Sec.~\ref{sec:coQCMIproperties} imply that co(R\'enyi entropy) is a monotonically non-increasing function of the decoherence rate. Given our numerical observations, it is then natural to wonder if the decomposition in Eq.~\eqref{eq:optimaldecompose} is perhaps optimal also for other quantities, including  co(R\'enyi entropy) and co(von Neumann entropy) (= entanglement of formation), so that the bipartite co(R\'enyi entropy) of the decohered mixed state equals the bipartite R\'enyi entropy of the pure state $|\psi(\beta)\rangle$ depicted in Fig.~\ref{fig:result_S}.

\section{Discussion} \label{sec:discussion}
In this paper, we introduced a measure of long-range entanglement in mixed states (abbreviated as co(QCMI)), given by the minimum value of the average TEE of the density matrix over all possible pure state decompositions (Eq.~\eqref{eq:def_coQCMI}). By construction, it is zero if and only if the density matrix admits a decomposition in terms of pure states with zero TEE. Furthermore, whenever a density matrix $\rho_2$ is obtained from a density matrix $\rho_1$ via a quantum channel that has a representation in terms of Kraus operators that are products of onsite unitaries, then co(QCMI)$[\rho_2] \leq $ co(QCMI)$[\rho_1]$. We focused on salient features of co(QCMI) in the context of decohered topological states, especially toric code subjected to bit-flip or phase-flip noise. We showed that below the error-recovery threshold, the density matrix cannot be written as a convex sum of SRE states, and relatedly, that co(QCMI) goes from non-zero to zero across the transition. These arguments apply more broadly to other topological ordered in general dimensions. For the 2d toric code, we then provided analytical and numerical support for the conjecture that the co(QCMI) equals TEE of a specific pure state that was recently introduced in Ref.~\cite{Chen_Separability_2024}. In particular, we developed a tensor-assisted Monte Carlo (TMC) algorithm to study the second R\'enyi TEE of the aforementioned pure state and found it satisfies the constraints that co(QCMI) must satisfy, thereby providing a non-trivial consistency check for our conjecture. We also numerically studied the scaling of the anyon condensation order parameter close to the transition, and found that the results match quite well with the known exponents of the RBIM along the Nishimori line. We anticipate that an analogous relation between co(QCMI) and TEE of a pure state  (Eq.\ref{eq:conjecture}) will hold true also for other examples discussed in Ref.\cite{Chen_Separability_2024}, e.g., 3d toric code or fracton states subjected to bit-flip or phase-flip noise. Our main results are also summarized in the last paragraph of Sec.~\ref{sec:intro}. 

There are currently several perspectives on mixed-state phase transitions in topological systems \cite{dennis2002,wang2003confinement, lee2023quantum, fan2023diagnostics,bao2023mixed,wang2023intrinsic,Chen_Separability_2024,sang2023mixed,sang2024stability,li2024replica,su2024tapestry, lee2024exact,lyons2024understanding, sohal2024noisy,ellison2024towards,lu2024disentangling, kikuchi2024anyon}. Our work connects at least two of these: one based on mixed-state phase equivalence using local, finite-time Lindbladian evolution \cite{coser2019classification,sang2023mixed,sang2024stability}, and another focused on long-range entanglement/separability \cite{Chen_Separability_2024,chen2024symmetryenforced}. Specifically, we showed that if the density matrix admits a decomposition in terms of short-range entangled pure states (in other words, if the density matrix is short-range entangled \cite{werner1989,hastings2011topological}), then the mixed-state cannot be connected to the pure topological state via a low-depth local channel (Sec.\ref{sec:coQCMIproperties}). It will be interesting to relate separability/entanglement to other perspectives such as coherent information \cite{fan2023diagnostics,lee2024exact}.

Let us  discuss potential challenges with the practical utility of co(QCMI). Perhaps the most formidable one is that calculating co(QCMI) for generic density matrices is rather difficult since it requires optimization over all possible decompositions of the density matrix in terms of pure states. One perspective one may take is that even if one can't calculate co(QCMI) for a given density matrix, one may be able to  put bounds on it by considering suitable decompositions of the density matrix. Combined with the general properties of the co(QCMI) (e.g., positivity and monotonicity), one may then use these bounds to constrain global aspects of the phase diagrams. This is indeed the route we took in this paper for mixed states obtained by locally decohering a topological state. Similar ideas may also be helpful for characterizing topological order in Gibbs states of topologically ordered systems. For example, the Gibbs state of 2d and 3d toric code at any non-zero temperature may be explicitly written as a convex sum of states that are SRE \cite{lu2020detecting}, and  therefore one expects that co(QCMI) vanishes at any non-zero $T$. For 4d toric code \cite{dennis2002}, on the other hand, one obtains a bound that co(QCMI) $\leq \log(2)$ below the finite-T quantum memory phase transition. Using similar arguments to those for the 2d toric code under local decoherence, one may argue that this bound is saturated. One may also consider a more ambitious approach of using numerical optimization methods to estimate co(QCMI), similar to the ones that have been used to estimate entanglement of formation (see, e.g., Ref.~\cite{audenaert2001variational}). Finally, as discussed in Sec.~\ref{sec:coQCMIproperties}, sometimes one may be able to exploit symmetries to calculate the co(QCMI) (or at least make an educated guess).

The second potential issue with co(QCMI) is that akin to pure state TEE, zero co(QCMI) is neither a sufficient nor a necessary condition for a state to be SRE (recall we define an SRE state is one that can be created via poly(log) depth circuit). It is not a sufficient condition because the mixed state may admit a decomposition in terms of GHZ-entangled pure states that have zero Levin-Wen TEE, but non-zero mutual information between distant subregions. \change{One way to characterize such states is to also calculate their co(MI) between distant subregions, which, unlike co(QCMI), will be sensitive to long-range entanglement encoded in GHZ-type states (and more generally, entanglement that can be captured by few point correlation functions).} A more interesting possibility is that the mixed-state admits a decomposition in terms of pure states that all have zero Levin-Wen TEE as well as exponentially decaying mutual information, but which are not ground states of a gapped, local Hamiltonian. Such pure states are not guaranteed to be SRE \cite{zhang2024unpublished} (or at least we do not know of a proof that shows to the contrary). Zero co(QCMI) is not a necessary condition for a state to be SRE due to the possibility of spurious TEE \cite{bravyi2008spurious,cano2015interactions,liujun2016spurious,williamson2019spurious}. As discussed in Sec.\ref{sec:coQCMIproperties}, this can be remedied by introducing a modified version of co(QCMI), see Eq.~\ref{eq:def_coQCMIv2}. Despite these potential drawbacks, it seems fair to say that QCMI (i.e. Levin-Wen TEE) in a pure state captures at least one kind of multi-partite entanglement that is a hallmark of known topologically ordered phases, and it is also non-zero for known generic, gapless ground states such as those corresponding to CFTs or compressible matter such as Fermi liquids. Therefore, if QCMI in a pure state vanishes, it is not unreasonable to say that the state has less long-range entanglement in a literal sense compared to a state with non-zero QCMI, even if the state with zero QCMI happens to have a large circuit complexity (see, e.g., recent discussion, Ref.\cite{li2024much}, distinguishing long-range entanglement in a GHZ state from that in a topologically ordered state, using maximum overlap between the state under consideration and a short-ranged entangled state).

On a related note, one may also define ``co(complexity)'' of a mixed state: 
     \be 
      \textrm{co(complexity)}[\rho] = \textrm{inf} \{\sum_i p_i \,\mathcal{C}(|\psi_i\rangle)\} \label{eq:def_cocomplexity}
     \ee 
      where $\mathcal{C}(|\psi_i\rangle)$ is the circuit complexity of the pure state $|\psi_i\rangle$, and the infimum is again taken over all possible decompositions of the mixed state $\rho$ as $\rho = \sum_i p_i |\psi_i\rangle \langle \psi_i|$. Recall that a circuit complexity of a pure state is the minimum depth of the circuit (which is assumed to be made of geometrically local, finite range gates) required to prepare it. co(complexity) was originally introduced in Ref.~\cite{agon2019subsystem} where it was called `ensemble complexity'. Let us  consider a mixed state $\rho_2$ that is obtained from a mixed state $\rho_1$ via a low-depth local channel that can be represented in terms of unitary Kraus operators. Following the same argument as in Sec.~\ref{sec:coQCMIproperties}, then the asymptotic scaling of the co(complexity) of a mixed state (with respect to the total system size) cannot increase under such a channel. For example, if the original mixed state has a co(complexity) of order $L^{\alpha}$, then the co(complexity) of the post-channel mixed state can't scale faster than $ L^{\alpha}$. One advantage of co(complexity) is that there is no analog of `spurious complexity' for obvious reasons, and hence, in this sense, it is a more robust quantity than TEE or co(QCMI). The challenge of course is that it seems extremely hard to calculate, since it requires two levels of optimizations, one over all pure state decompositions, and the other over all possible circuits for each $|\psi_i\rangle$ in  a specific decomposition. 

Another aspect that needs more thought is the choice of the tetra-partition used to define the co(QCMI). To obtain co(QCMI), one needs to minimize the average TEE over all possible pure state decompositions of the density matrix, including the ones that are not translationally invariant. It is then not obvious if co(QCMI) is independent of the partition used to define it. Should one average it over all possible tetra-partitions, or take the minimum over all possible tetra-partitions? Similar questions can also be raised for the entanglement of formation as a measure of bipartite mixed-state entanglement, or even pure state TEE in a non-translationally invariant system. We are not aware of any detailed discussion of such questions in the literature. 

Finally, we note that the TMC method developed in this work is likely to have several more applications in the context of 2d mixed states. Local decoherence of 2d quantum systems naturally leads to wavefunctions whose amplitudes are related to 2d classical statistical mechanics models, and therefore, it will be expedient to apply the TMC method to these problems, such as calculating the R\'enyi negativity across mixed-state phase transitions, or the study of critical pure states that are related to decohered mixed states. It will be also worthwhile to improve the TMC method along the lines for other models ~\cite{liaoControllable2023,zhangIntegral2024,zhouIncremental2024} so that the R\'enyi TEE can be calculated with polynomial complexity in system size for a fixed relative error.

\textbf{\underline{Note added}}: While this work was being completed, we became aware of an upcoming work, Ref.\cite{lessa2025higher}, whose authors have also independently studied mixed-state entanglement defined via the convex-roof construction of QCMI, and its general properties that overlap with our work.

\section*{Acknowledgment}
We thank Subhayan Sahu, Tim Hsieh, and especially Tsung-Cheng Lu for helpful discussions, and Bin-Bin Chen and Yu-Hsueh Chen for valuable discussions on the tensor network implementation. We also thank Bowen Shi, Chao-Ming Jian, John McGreevy and Ruben Verresen for valuable feedback on the manuscript, and John Chalker for a very helpful correspondence on Ref.\cite{merz2002negative}. TTW, MHS and ZYM acknowledge the support from the Research Grants Council (RGC) of Hong Kong Special Administrative Region of China (Project Nos. 17301721, AoE/P-701/20, 17309822, HKU C7037-
22GF, 17302223), the ANR/RGC Joint Research Scheme sponsored by RGC of Hong Kong and French National Research Agency (Project No. A HKU703/22), the GD-NSF (No. 2022A1515011007). We thank HPC2021 system under the Information Technology Services and the Blackbody HPC system at the Department of Physics, University of Hong Kong, as well as the Beijing PARATERA Tech CO.,Ltd. (URL: https://cloud.paratera.com) for providing HPC resources that have contributed to the research results reported within this paper. TG is supported by the National Science
Foundation under Grant No. DMR-1752417. We acknowledge the hospitality of Kavli Institute for Theoretical Physics (KITP) and thank the organizers of the KITP program ``Correlated Gapless Quantum Matter''  where this work was initiated. This research was supported in part by grant NSF PHY-2309135 to the KITP.

\appendix

\section{Proof that mixed state for $p < p_c$ is not a convex sum of SRE pure states} \label{sec:notSRE}

\change{In this appendix we will prove Theorem~\ref{prop:longrange} and associated Corollary~\ref{corollary:longrange}. In particular, we will show that the density matrix of a CSS topological code under the action of local decoherence cannot be written as a convex sum of SRE pure states for $p < p_c$}, i.e., the density matrix is long-range entangled in the mixed-state phase where error-correction works. The main idea is to combine the following four constraints:

\begin{enumerate}
    \item  If two mixed states are in the same phase of matter, then there exists a low-depth local quantum channel that connects them in either direction Refs.\cite{coser2019classification,sang2023mixed,sang2024stability}.

        \item A low-depth local channel acting on a pure SRE state results in a density matrix whose connected correlations are short-ranged. This follows from Lieb-Robinson bound \cite{lieb1972finite,poulin2010liebrobinson}. To see this explicitly, we recall that a low-depth local quantum channel acting on a pure state $|\textrm{SRE}\rangle$ of the system is equivalent to applying a low-depth local unitary $U$ on $|\textrm{SRE}\rangle \otimes |0\rangle_a$ where $|0\rangle_a$ denotes the product state of ancillae, followed by tracing out ancillae. Consider the connected correlation function $C(x,y) = \langle O_1(x) O_2(y) \rangle - \langle O_1(x) \rangle \langle O_2(y) \rangle$ with respect to the state $U|\textrm{SRE}\rangle \otimes |0\rangle_a$, where $O_1, O_2$ are operators that live in the Hilbert space of the system. $C(x,y)$ also equals the connected correlation function $\langle \tilde{O_1}(x) \tilde{O_2}(y) \rangle - \langle \tilde{O_1}(x) \rangle \langle \tilde{O_2}(y) \rangle$ with respect to the SRE state $|\textrm{SRE}\rangle \otimes |0\rangle_a$, where $\tilde{O} = U^{\dagger} O U$. As long as $|x-y|$ is much bigger than the depth of the unitary $U$, operators $\tilde{O_1}(x)$ and $\tilde{O_2}(y)$ do not overlap, and therefore $C(x,y)$ decays exponentially, since 
        $|\textrm{SRE}\rangle \otimes |0\rangle_a$ is SRE.
    
    \item  Topological ordered pure states have long-range correlations for logical operators supported on non-contractible regions \cite{jian2015long}.

    \item If two Hermitian operators $O_1, O_2$ that satisfy $O^2_1 = O^2_2 = \mathds{1}$ mutually anti-commute, then their expectation value with respect to any pure state $|\psi\rangle$ satisfies $\langle \psi|O_1|\psi\rangle^2 + \langle \psi|O_2|\psi\rangle^2 \leq 1$ (see, e.g., Ref.\cite{toth2005entanglement}).
\end{enumerate}

Let's assume that for $p < p_c$, the density matrix $\rho(p)$ admits a decomposition in terms of SRE pure states, i.e., $\rho(p) = \sum_a \, p_a |\textrm{SRE}_a\rangle \langle \textrm{SRE}_a|$ where $p_a$ (not to be confused with $p$, the decoherence rate) is the probability for the state $|\textrm{SRE}_a\rangle $. The first constraint listed above implies \cite{coser2019classification,sang2023mixed,sang2024stability} that for $p < p_c$ there exists a constant time quasi-local Lindblad evolution $\mathcal{L}(\tau)$ that approximately converts the mixed state $\rho(p)$ to the pure toric code ground state $\rho(p=0)$. That is, 

\be \big|\mathcal{T} e^{\int_{0}^{1} \, dt\, \mathcal{L}(t)} \rho(p) - \rho(p=0)\big|_1 \leq \epsilon \label{eq:tracenorm}
\ee 
where $\mathcal{T}$ denotes time-ordering, $|\cdot \big|_1$ denotes the trace norm and $\epsilon$ is the tolerance that can be taken to vanish as $1/\textrm{poly}(L)$ where $L$ is the total system's linear length. We will now show that the constraint $\rho(p) = \sum_a \, p_a |\textrm{SRE}_a\rangle \langle \textrm{SRE}_a|$  implies that $\epsilon \geq (3-\sqrt{5})/2 \approx 0.38$, which is a contradiction with the requirement that $\epsilon$ can be taken arbitrarily small for $p < p_c$ \cite{sang2023mixed,sang2024stability}. Therefore, the assumption $\rho(p) = \sum_a \, p_a |\textrm{SRE}_a\rangle \langle \textrm{SRE}_a|$ must be incorrect.

Let us write the action of Lindblad evolution on a particular pure state $|\textrm{SRE}_a\rangle$ that enters the convex decomposition of $\rho(p)$ as

\be 
\mathcal{T} e^{\int_{0}^{1} \, dt\, \mathcal{L}(t)} \left(|\textrm{SRE}_a\rangle  \langle \textrm{SRE}_a|\right) = \sum_m q_{a,m} |\phi_{a,m}\rangle \langle \phi_{a,m}| \label{eq:rho_a}
\ee 
Note that the decomposition on the r.h.s. in the above equation is not unique, and the following discussion is independent of which particular decomposition is chosen.

To obtain the aforementioned bound on $\epsilon$, we will consider expectation values of operators made out of three distinct logical operators $\overline{X}, \overline{Z}^A,\overline{Z}^B$ (see Fig.\ref{fig:logical_strings}) \change{in the underlying CSS topological code. The logical operator $\overline{X}$ is conjugate to both $\overline{Z}^A,\overline{Z}^B$ (i.e. it has a non-zero intersection number with $\overline{Z}^A,\overline{Z}^B$), and therefore satisfies $\overline{X} \overline{Z}^A = - \overline{Z}^A \overline{X}$, and $\overline{X} \overline{Z}^B = - \overline{Z}^B \overline{X}$.} We will choose $\rho(p=0)$ as the toric code ground state that is an eigenstate of $\overline{X}$ with eigenvalue 1. This implies that $ \tr(\rho(p=0) \overline{X}) = 1$, $ \tr(\rho(p=0) \overline{Z}^A \overline{Z}^B) = 1$, and  $ \tr(\rho(p=0) \overline{Z}^A) = \tr(\rho(p=0) \overline{Z}^B) = 0$. 

Using the second constraint above, the connected correlation function $\langle  \overline{Z}^A  \overline{Z}^B \rangle - \langle  \overline{Z}^A \rangle  \langle  \overline{Z}^B \rangle  $ with respect to the state 
$\mathcal{T} e^{\int_{0}^{1} \, dt\, \mathcal{L}(t)} \left(|\textrm{SRE}_a\rangle  \langle \textrm{SRE}_a|\right)$ decays exponentially. Therefore, upto exponentially small corrections in the total system size that we can safely neglect (we are interested in the thermodynamic limit), one finds, for each `$a$' separately,

\bea 
& & \sum_m  q_{a,m} \langle \phi_{a,m}|  \overline{Z}^A  \overline{Z}^B| \phi_{a,m}\rangle  \nonumber \\
 &  & = \hspace{-0.1cm}\sum_{m,m'}  q_{a,m}  q_{a,m'} \langle \phi_{a,m}|  \overline{Z}^A| \phi_{a,m}\rangle \langle \phi_{a,m'}| \overline{Z}^B| \phi_{a,m'}\rangle \nonumber \\
 & & =  \sum_{m,m'}  q_{a,m}  q_{a,m'}z^A_{a,m} z^B_{a,m'} \label{eq:ZAZB}
\eea 
where $z^A_{a,m} =  \langle \phi_{a,m}|  \overline{Z}^A| \phi_{a,m}\rangle$ and similarly $z^B_{a,m} =  \langle \phi_{a,m}|  \overline{Z}^B| \phi_{a,m}\rangle$.

Let us consider the consequence of Eq.\ref{eq:tracenorm} for the density matrix $\mathcal{T} e^{\int_{0}^{1} \, dt\, \mathcal{L}(t)} \rho(p) = \sum_a p_a q_{a,m} |\phi_{a,m}\rangle \langle \phi_{a,m}|$. The trace-norm distance between two density matrices bounds the difference in expectation value of all operators whose eigenvalues lie between 0 and 1. 
Eq.\ref{eq:tracenorm}, along with Eq.\ref{eq:ZAZB}, then implies

\bea
\sum_{a,m} p_a q_{a,m} x_{a,m} \geq 1-\epsilon \nonumber \\
\sum_{a,m,m'} p_a  q_{a,m}  q_{a,m'}z^A_{a,m} z^B_{a,m'} \geq 1-\epsilon \label{eq:implytracenorm}
\eea 
where $x_{a,m} =  \langle \phi_{a,m}|  \overline{X}| \phi_{a,m}\rangle$. The first of these equations follows from comparing the expectation value of $\overline{X}$ with respect to the states $\mathcal{T} e^{\int_{0}^{1} \, dt\, \mathcal{L}(t)} \rho(p)$ and $\rho(p=0)$, while the second one follows from comparing the expectation value of $\overline{Z}^A \overline{Z}^B$ with respect to these two states (supplemented by Eq.\ref{eq:ZAZB}).

Finally, since  $\overline{X} \overline{Z}^A = - \overline{Z}^A \overline{X}$, and $\overline{X} \overline{Z}^B = - \overline{Z}^B \overline{X}$, and all three operators $\overline{X}, \overline{Z}^A, \overline{Z}^B$ square to identity, the fourth constraint above implies that

\bea 
\left(x_{a,m}\right)^2 + \left(z^A_{a,m}\right)^2 \leq 1 \nonumber \\
\left(x_{a,m}\right)^2 + \left(z^B_{a,m}\right)^2 \leq 1 \label{eq:uncertainity}
\eea 
 for any $a, m$. It is easy to see that Eqs.\ref{eq:implytracenorm} and \ref{eq:uncertainity} are inconsistent with each other when $\epsilon \ll 1$. Indeed, when $\epsilon$ exactly equals zero, Eqs.\ref{eq:implytracenorm} imply that $\left(x_{a,m}\right)^2 = \left(z^A_{a,m}\right)^2 = 1$ which is in clear contradiction with Eqs.\ref{eq:uncertainity} (recall that $p_a$ and $q_{a,m}$ are normalized probabilities, i.e., $\sum_a p_a = 1$ and for any $a$, $\sum_m q_{a,m} = 1$). To obtain a bound on $\epsilon$, we start with Eq.\ref{eq:implytracenorm} and  apply Cauchy-Schwarz inequality while using Eq.\ref{eq:uncertainity}:
 
 \bea 
& & (1 - \epsilon) \leq  \sum_{a,m,m'} p_a  q_{a,m}  q_{a,m'}z^A_{a,m} z^B_{a,m'} \nonumber \\
& & \leq \sum_{a,m,m'} p_a q_{a,m} q_{a,m'}\sqrt{1- x^2_{a,m}} \sqrt{1- x^2_{a,m'}}\nonumber \\
& & = \sum_a p_a \left(\sum_m q_{a,m} \sqrt{1- x^2_{a,m}}\right)^2 \nonumber \\
& & \leq \sum_a p_a q_{a,m} (1- x^2_{a,m}) \nonumber \\
& & \leq 1 - (1-\epsilon)^2 \label{eq:cauchy}
 \eea 
 where in the last sentence we have used Eq.\ref{eq:implytracenorm} as $\sum_{a,m} p_a q_{a,m} x^2_{a,m} \geq \left(\sum_{a,m} p_a q_{a,m} x_{a,m}\right)^2 \geq (1-\epsilon)^2$. Therefore, one obtains $(1-\epsilon) + (1-\epsilon)^2 \leq 1$, which can be satisfied only if $\epsilon \geq (3-\sqrt{5})/2 \approx 0.38$. This is incompatible with the requirement that error-recovery is possible for $p < p_c$, i.e., there exists a low-depth local channel that can take the mixed state back to the undecohered toric code ground state \cite{sang2023mixed,sang2024stability}.

 As mentioned in the main text, one may strengthen the above argument by allowing for a non-zero topological ordered component in the density matrix $\rho(p)$. In particular, let us consider the possibility that 

 \be 
\rho(p) = \sum_a \, p'_a |\textrm{SRE}_a\rangle \langle \textrm{SRE}_a| + (1- \sum_a p'_a) \rho(p=0) \ee 
where $\rho(p=0)$ is of course the pure toric code ground state (note that $\sum_a p'_a < 1$, and therefore $\{p'_a\}$ is not a normalized probability probability distribution). Repeating the same argument as above, the analog of Eqs.\ref{eq:implytracenorm} is:

\bea
\sum_{a,m} p'_a q_{a,m} x_{a,m} + (1-\sum_a p'_a) \geq 1-\epsilon \nonumber \\
\sum_{a,m,m'} p'_a  q_{a,m}  q_{a,m'}z^A_{a,m} z^B_{a,m'} + (1-\sum_a p'_a) \geq 1-\epsilon \nonumber
\eea 
where we have used the fact that the maximum value of the expectation values $ \tr\left(\mathcal{T} e^{\int_{0}^{1} \, dt\, \mathcal{L}(t)} \rho(p=0) \overline{X}\right)$ and $ \tr\left(\mathcal{T} e^{\int_{0}^{1} \, dt\, \mathcal{L}(t)} \rho(p=0) \overline{Z}^A\, \overline{Z}^B\right)$ is unity. The above equations may be rewritten as
\bea
\sum_{a,m} p_a q_{a,m} x_{a,m} \geq 1-\frac{\epsilon}{w_{\text{SRE}}} \nonumber \\
\sum_{a,m,m'} p_a  q_{a,m}  q_{a,m'}z^A_{a,m} z^B_{a,m'} \geq 1-\frac{\epsilon}{w_{\text{SRE}}} \label{eq:implytracenorm_modified}
\eea 
where $p_a = p'_a/\sum_b p'_b$ is the normalized probability distribution function, and $w_{\text{SRE}} = \sum_a p'_a$ is the total weight of SRE states in the density matrix $\rho(p)$.
Eqs.\ref{eq:uncertainity} remain unchanged. Therefore, the structure of the new equations is identical to the old ones, with the replacement $\epsilon \rightarrow \epsilon/w_{\text{SRE}}$. Therefore, using the same set of inequalities as before (Eqs.\ref{eq:cauchy}), one obtains the constraint

\be 
\frac{\epsilon}{w_{\text{SRE}}} \geq (3-\sqrt{5})/2 \approx 0.38.
\ee 
Therefore, as $\epsilon \rightarrow 0$, the total weight of the SRE states, ${w_{\text{SRE}}}$, also goes to zero.

\section{Analysis of the critical exponents} \label{sec:AC_CE}

\begin{figure}[htp!]
\centering
\includegraphics[width=\columnwidth]{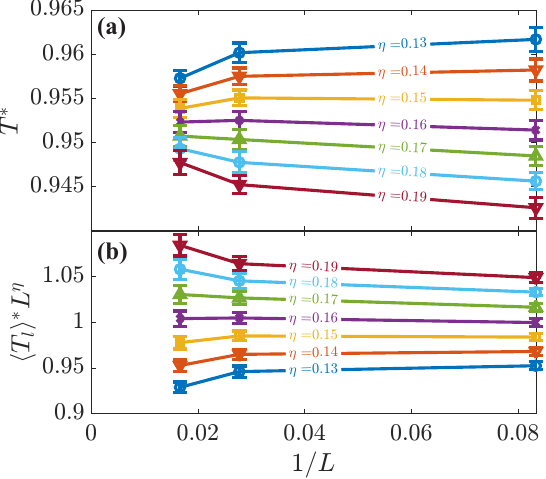}
\caption{\textbf{Crossing point analysis.} Scaling of the $x$ and $y$ coordinate of the crossing points against $1/L$. Each of them is the interception of the rescaled curves of $L$ and $2L$.}
\label{fig:AC_cross}
\end{figure}

\begin{figure}[t]
\centering
\includegraphics[width=\columnwidth]{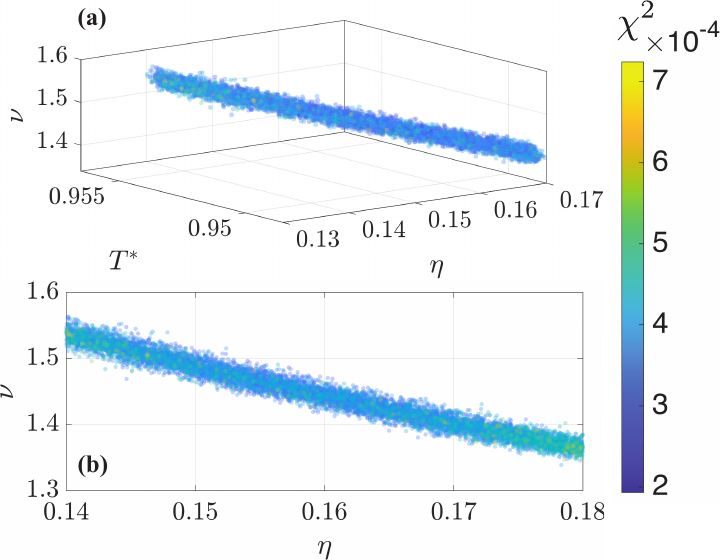}
\caption{\textbf{Quality of data collapse.} Result of $10^4$ number of minimization processes. The two panels show the final $\nu$ and $T_c$ with corresponding input $\eta$, and the value of the loss function $\chi^2$. Blue (yellow) dots indicate a collapse with a smaller (larger) $\chi^2$.}
\label{fig:AC_CE}
\end{figure}

We first try to find the crossing points of the rescaled average $\av{T_l}L^\eta$ between data obtain from $L$ and $2L$, and see their trend against $1/L$. And the $x$ ($y$) axis of the crossing point is denoted as $T^*$ ($\av{T_l}^*L^\eta$). 

As shown in Fig.~\ref{fig:AC_cross}, both $x$ and $y$ coordinate stay nearly at constants when one choose $\eta=0.16$. Indeed, the crossing point need not to be stay constant exactly, instead they can converge algebraically with power determined by the next scaling dimension in line. However, for cases with $\eta$ out of the range $0.16\pm 0.02$, the $y$ coordinate do not converge up to the largest system size, which suggests the anomalous dimension $\eta=0.16(2)$.

With one exponent and its error bar determined, we then try to collapse the data by scaling also the horizontal axis to $\mu=(T-T_c)L^\nu$, and minimizing the loss function $\chi^2$ by varying $T_c$ and $\nu$. The loss function is defined as
\be
\chi^2=\frac{S_\textrm{res}}{S_\textrm{tot}}=\frac{\sum_i(y_i-\hat{y}_i)^2}{\sum_i(y_i-\bar{y})^2},
\ee
where $y_i$ is the rescaled data $\av{T_l}L^\eta$, $\hat{y}_i$ is corresponding function value of a polynomial function fitted using $\mu$ and $y_i$ from all system sizes, and $\bar{y}$ is the mean value of $y_i$. A good set of critical point and exponents should be able to collapse all data points to a smooth curve, thus minimizes $S_{res}$ and $\chi^2$.

We repeat the minimization process for $10^4$ times by inputting $\eta$ choosing from the range $0.16\pm 0.02$, and the numerical result $\av{T_l}$ with perturbation within its error-bar to include also the statistical error.

Fig.~\ref{fig:AC_CE} shows the result of the minimization. With this window of $\eta$ chosen, the correlation length exponent varies within $1.44\pm 0.12$, and $T_c$ within $0.951\pm 0.005$. There are more blue dots (indicating lower $\chi^2$) and less yellow dots (indicating higher $\chi^2$) in the middle region, which indicates good estimation on the critical point and exponents.

\newpage
\input{main.bbl}

\end{document}

%% file: main.bbl
%